\documentclass[conference]{IEEEtran}
\IEEEoverridecommandlockouts
\usepackage{amsmath,amsfonts}
\usepackage{algorithmic}
\usepackage{array}
\usepackage[caption=false,font=scriptsize,labelfont=sf,textfont=sf]{subfig}
\usepackage{textcomp}
\usepackage{stfloats}
\usepackage{url}
\usepackage{verbatim}
\usepackage{graphicx}
\usepackage{amsmath,amssymb,amsfonts}
\usepackage{algorithmic}
\usepackage{algorithm}
\usepackage{makecell}
\usepackage{multirow}

\def\BibTeX{{\rm B\kern-.05em{\sc i\kern-.025em b}\kern-.08em
    T\kern-.1667em\lower.7ex\hbox{E}\kern-.125emX}}
\usepackage{balance}
\usepackage{hyperref}
\usepackage{xcolor}
\hypersetup{hidelinks,
	colorlinks=true,
	citecolor = green,
	urlcolor = magenta,
	pdfstartview=Fit,
	breaklinks=true}
\usepackage[numbers, sort]{natbib}
\begin{document}
\title{Unleashing the Power of Self-Supervised Image Denoising: A Comprehensive Review\\
}

\author{Dan Zhang*, 
\IEEEmembership{Member, IEEE}, Fangfang Zhou*, Felix Albu
\IEEEmembership{Senior Member, IEEE}, \\
Yuanzhou Wei, Xiao Yang, Yuan Gu, Qiang Li
\thanks{*These authors contributed to the work equllly and should be regarded as co-first authors. Email: zhangdan\underline{ }fiona@163.com, 51174700059@stu.ecnu.edu.cn, felix.albu@gmail.com, ywei011@fiu.edu, xy50573@uga.edu, uwin@gwu.edu, qiang.li@rwth-aachen.de.}}

\markboth{Journal of \LaTeX\ Class Files,~Vol.~18, No.~9, September~2020}%
{How to Use the IEEEtran \LaTeX \ Templates}

\maketitle

\begin{abstract}
The advent of deep learning has brought a revolutionary transformation to image denoising techniques. However, the persistent challenge of acquiring noisy-clean pairs for supervised methods in real-world scenarios remains formidable, necessitating the exploration of more practical self-supervised image denoising. This paper focuses on self-supervised image denoising methods that offer effective solutions to address this challenge. Our comprehensive review analyzes the latest advancements in self-supervised image denoising approaches, categorizing them into three distinct classes: General methods, Blind Spot Network (BSN)-based methods, and Transformer-based methods. For each class, we provide a concise theoretical analysis along with their practical applications. To assess the effectiveness of these methods, we present both quantitative and qualitative experimental results on various datasets. Additionally, we critically discuss the current limitations of these methods and propose promising directions for future research. By offering a detailed overview of recent developments in self-supervised image denoising, this review serves as an invaluable resource for researchers and practitioners in the field, facilitating a deeper understanding of this emerging domain and inspiring further advancements.
\end{abstract}


\section{Introduction}
\IEEEPARstart{N}{oise} in images often manifests as isolated pixels or pixel blocks that have a significant visual impact, disrupting the actual information content of the image and making it unclear. Generally, noise signals are unrelated to the object being imaged and represent useless information. Noise arises from two main sources: during image acquisition and during image transmission and processing. Image sensors like Charge-coupled Device (CCD) and Complementary Metal Oxide Semiconductor (CMOS) arrays introduce noise due to sensor material properties, the environment, electronic components, and circuitry.  This includes thermal noise from resistance, transistor channel noise, photon noise, dark current noise, and non-uniform light response noise \cite{xu2010new, li2015low}. Digital images become contaminated with various types of noise \cite{henkelman1985measurement, wang2021electronic} during transmission and recording due to imperfect media and equipment. Even slight camera shake during shooting can generate noise \cite{liu2022shaking, komatsu2019effectiveness}. Noise can also be introduced at different image processing stages when inputs diverge from expectations. Image noise typically exhibits the following characteristics: It is irregular and random in distribution and magnitude, and generally correlated with the image \cite{boncelet2009image, zhao2014robust}.  
There are several types of noise based on its probability distributions that can affect images: Gaussian noise \cite{slepian1962one}, Poisson noise \cite{middleton1951theory}, multiplicative noise \cite{mcintyre1966multiplication}, salt-and-pepper noise \cite{cockayne1936dwarfism}, Gamma noise \cite{schultz1964shutdown}, Rayleigh noise \cite{ronken1969intensity}, Uniform noise \cite{eckart1953theory} and Exponential noise \cite{jerde1967effects}.

\begin{figure}[!t]
	\setlength{\abovecaptionskip}{0cm}
	\centering
	\includegraphics[width=0.98\linewidth]{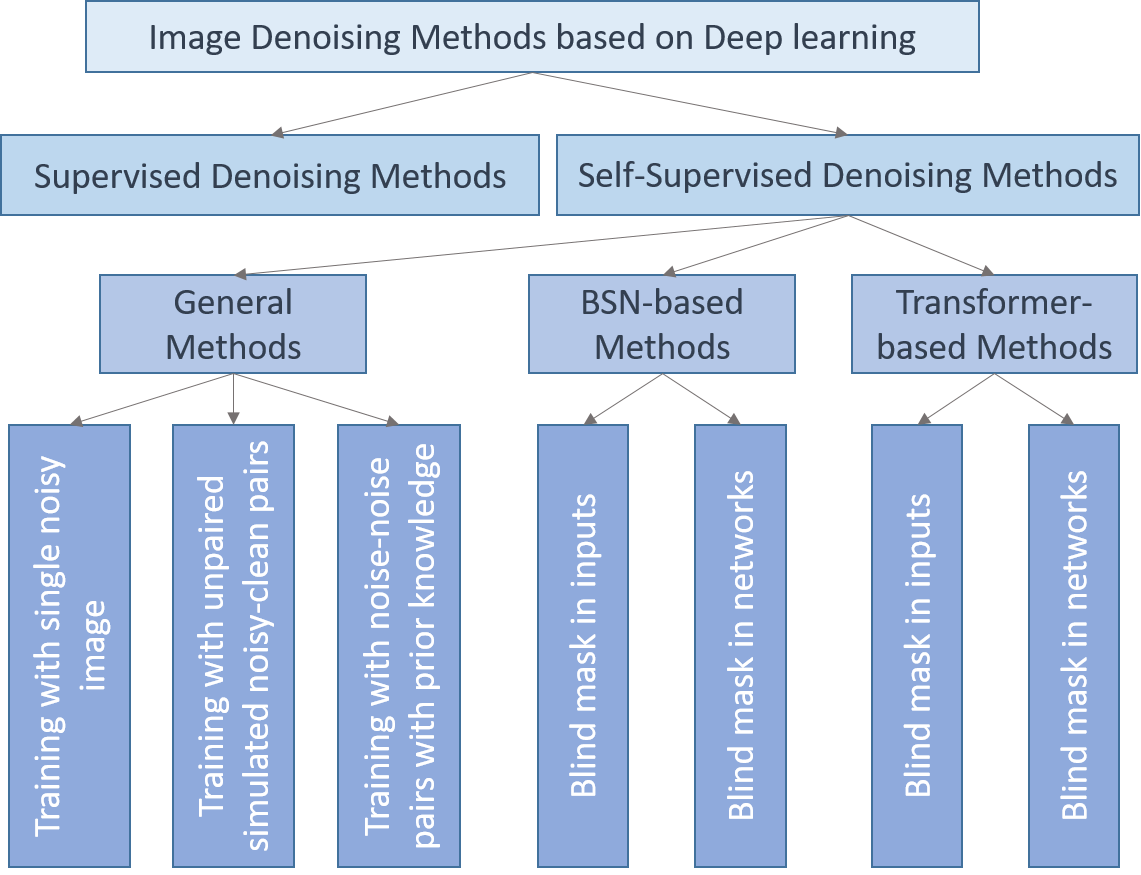}
	\caption{Outline of the survey.}
	\label{fig:1}
\vspace{-2em}
\end{figure}

Image noise is a common problem that can degrade the visual quality of images, making it difficult to analyze and negatively impacting subsequent tasks such as image detection \cite{yang2022deep,han2023ssgd, zhou2020automatically}, classification \cite{lu2023deep, xiang2022tkil,wei2023breast}, segmentation \cite{wang2022adversarial, huang2022simultaneous, zhao2022automated, you2022incremental}, tracking \cite{lee2017online, tang2019moana}, text generation \cite{luo2017text, luo2016text} and more. Therefore, image denoising methods for restoring a clean image $x$ from a noisy image $y$ = $x$ + $n$ is necessary for improving the visual quality of images and facilitating subsequent image analysis tasks.

Traditional image denoising methods \cite{dabov2007image}, \cite{buades2011non, buades2005non, starck2002curvelet, dabov2006image, mihcak1999low, chen1999tri, zhang2008multiresolution} work reasonably well for their designed/tuned scenarios but do not generalize or scale to achieve human-level denoising performance across the range of noise types, levels, complexities, data sizes, and quality metrics required for most real-world applications. Progress in denoising now depends on machine learning techniques that can overcome these challenges. Sparse-based methods is first applied to image denoising such as \cite{dabov2007image}, non-locally centralized sparse representation (NCSR) method \cite{dong2012nonlocally}. Then, in order to save computer costs and recover more image details, more efficient methods were proposed \cite{elad2006image, schmidt2014shrinkage, osher2005iterative, ren2019simultaneous, zuo2014gradient, gu2014weighted}. Examples include Markov random field (MRF) \cite{schmidt2014shrinkage}, gradient models \cite{zuo2014gradient}, and weighted nuclear norm minimization (WNNM) \cite{gu2014weighted}. However, these models usually require manual hyperparameter setting, which is very unfriendly to different datasets that need to manually try to adjust parameters. Moreover, such traditional methods require a lot of computing resources in the denoising process.

As we enter the era of big data and experience significant improvements in computer computing power, various fields are leaning towards learning-based algorithms to meet future demands \cite{wei2023research,chen2022relax,wang2022adversarial,pokle2022contrasting, wang2022cnn,yang2022deep,mei2023mac,chen2020item,gong2016research,he2023hiercat,wang2023b,hong2020end,lu2023deep, luo2023efficient,gu2023unveiling}. Deep learning, in particular, has demonstrated superior performance and computational cost compared to traditional algorithms in many fields \cite{dabov2007image, wen2021openmem, wang2022simd, dabov2007image, minaee2021deep, tabernik2019deep, luo2015adaptive, gu2014weighted, goossens2009removal, makinen2019exact, jiang2023magnetic, sachdev2019machine, yang2022deep, bist2023mislaying, li2022atrous, ma2020statistical, ma2022traffic, he2021interpretable, lin2022cascade, wang2022cnn, alzubaidi2023survey}.
In the area of image denoising, deep learning methods can be categorized as either supervised or self-supervised, depending on whether or not they require noisy-clean image pairs.

{\bf Supervised image denoising} involves training a deep neural network (DNN) using pairs of noisy and corresponding clean images, and let the model learn the mapping of a noisy image to a clean image \cite{zhang2018ffdnet, guo2019toward, kim2020transfer, liu2019multi, park2019densely, chen2022simple, yue2020dual, xie2012image} or let the residual of a clean image and a noisy image be used as a target, and learn to separate the noise from the noisy image \cite{zhang2017beyond, anwar2019real, chang2020spatial}. Once trained, the model can take in a new noisy image and produce a denoised version of it. Jain et al. \cite{jain2008natural} was the first to use DNN for image denoising in 2008, and achieved denoising results comparable to the best traditional algorithms at the time, wavelet and MRF methods\cite{li1994markov, chen2007spatio}, while having a smaller computational cost. Then more DNN methods were proposed \cite{zhang2018ffdnet, vaksman2021patch, xie2012image, dong2018denoising, agostinelli2013adaptive}. Generally, the higher the depth and width of the network, the better the performance of the model, and correspondingly, the larger the number of parameters of the model. Due to the limitations of the computer memory at that time, these models require a trade-off between performance and the number of model parameters. Because actual noisy images can be readily acquired, whereas their corresponding clean images are less accessible, this leads to the challenge of obtaining perfect pairs of noisy-clean images. Therefore, in many studies, people use different noise adding strategies to obtain injected noise images based on limited clean datasets, and obtain artificially synthesized noisy-clean image pairs.

In addition to the commonly used additive white noisy images (AWNI), the noise with Poisson-Gaussian distribution can also be adopted as noise modelling. Zhang et al. \cite{zhang2017beyond} proposed the denoising convolutional neural network (DnCNN) in 2017 to train an image denoising model by manually adding Additive White Gaussian Noise (AWGN) on clean images to generate noisy-clean pairs. Through a residual learning, the model not only improves the model denoising performance, but also greatly reduces the amount of calculation. In \cite{guo2019toward}, Guo et al. proposed a convolutional blind denoising network (CBDNet) designed specifically for real photographs. The CBDNet utilizes the Poisson-Gaussian noise distribution and in-camera processing. It features two subnetworks: the noise estimation subnetwork and the non-blind denoising subnetwork. To enhance the robustness of the denoiser, the noise estimation subnetwork applies an asymmetric loss, with greater penalties for underestimating the noise level. Following the CBDNet's framework, a simpler yet highly effective blind denoising model called the SDNet was introduced in \cite{zhao2019simple}. The SDNet employs the generalized signal-dependent noise model \cite{liu2014practical}. By incorporating a stage-wise procedure and lifted residual learning, the SDNet achieves competitive results on both synthesized and real noisy images.

Deep learning methods are data-driven and rely heavily on the quality and quantity of training data for achieving high-quality denoising results. Consequently, training datasets are critical for these models. However, acquiring absolutely clean images is often impractical in real-world applications, making methods that do not require clean images even more valuable for research.  

{\bf Self-supervised image denoising} is an image denoising algorithm that does not require clean images. It has attracted a large number of researchers to study this field \cite{xu2020noisy, xie2020noise2same, quan2020self2self, huang2021neighbor2neighbor, zhang2022idr, kim2022noise, wang2022blind2unblind, neshatavar2022cvf, wang2023lg, zhang2023mm, zhang2023self, vaksman2023patch, li2023spatially, chen2018image,lehtinen2018noise2noise,krull2019noise2void,batson2019noise2self,laine2019high,wu2020unpaired,moran2020noisier2noise,krull2020probabilistic,pang2021recorrupted,kim2021noise2score,lee2022ap, el2020blind, xu2017patch}. Noise2Noise \cite{lehtinen2018noise2noise} was the first self-supervised algorithm that achieved performance comparable to supervised image denoising by training on only aligned noisy-noisy image pairs. However, obtaining a large number of two perfectly aligned noisy-noisy image pairs can be challenging. To address this issue, some researchers \cite{wu2020unpaired, moran2020noisier2noise, krull2020probabilistic, xu2020noisy, zhang2022idr} use noisy images and noise models to artificially generate noisy-noisy pairs for training. However, in practical scenarios, noise level estimates and priors may not be readily available. Some researchers have developed denoising models using a single noisy image for training \cite{krull2019noise2void, pang2021recorrupted, huang2021neighbor2neighbor, neshatavar2022cvf}. One notable example is the Blind Spot Network (BSN) proposed in \cite{krull2019noise2void}, and a large number of BSN-based methods \cite{krull2019noise2void, batson2019noise2self, wu2020unpaired, krull2020probabilistic, lee2022ap, honzatko2020efficient} have been proposed and have demonstrated superior denoising performance. In order to combine the advantages of convolutional neural network (CNN) in local feature extraction and Transformer in global feature extraction, some Transformer-based methods \cite{wang2023lg, zhang2023self, papkov2023swinia} have been proposed one after another, and have achieved excellent results in self supervised image denoising.

Although there have been several reviews on image denoising\cite{tian2020deep,ilesanmi2021methods, izadi2022image,kong2023comparison}, to the best of our knowledge, there is no dedicated survey of self-supervised image denoising algorithms, which makes it difficult for researchers to quickly understand the classic self-supervised image denoising algorithms and provide feasible directions for future research.This article aims to address this gap by investigating self-supervised image denoising algorithms in recent years. The main contributions of this paper are as following:

{\bf Firstly}, we propose a survey focusing solely on self-supervised image denoising methods and divide the involved methods into three categories: General methods, BSN-based methods and Transformer-based methods as shown in Fig. \ref{fig:1}.

{\bf Secondly}, we briefly introduce the denoising principles of representative methods in each category. 

{\bf Thirdly}, through the statistical analysis of a large number of experimental results, we evaluate the performance of different methods on different denoising tasks, discuss the existing challenges in image denoising, and provide suggestions for the future development direction of this field.

The organizational framework of this review is as follows: Section \ref{sec:Section2} divides self-supervised image denoising algorithms into General method, BSN-based methods and Transformer-based methods, and provides a brief introduction to the principles of the classic algorithms in each category, as well as statistics on applicable scenarios, while Section \ref{sec:Section3} introduces common image denoising datasets and evaluation metrics. Section \ref{sec:Section4} quantitatively and qualitatively analyzes the state-of-the-art self-supervised denoising methods in recent years. Section \ref{sec:Section5} discusses current challenges and future research directions for self-supervised denoising methods. Finally, Section \ref{sec:Section6} concludes the survey.

\section{Self-supervised image denoising methods}
\label{sec:Section2}
\noindent Self-supervised image denoising algorithm is a deep learning image denoising algorithm that does not require paired noisy-clean images as training data.

{\bf BSN} is a self-supervised image denoising algorithm based on the assumption that the noise is spatially independent and zero-mean \cite{krull2019noise2void}, and uses the spatial correlation of the image signal to predict blind pixels through surrounding pixels. 

In recent years, a large number of BSN-based image denoising algorithms have emerged \cite{krull2019noise2void,batson2019noise2self,laine2019high,wu2020unpaired, krull2020probabilistic, lee2022ap, honzatko2020efficient}, and Transformer has been proved to have irreplaceable advantages of CNN in the  extraction of global information in image processing tasks \cite{han2022survey, liu2021swin, zhao2022hybrid, khan2022transformers, zhou2023high, wang2023dealing, papkov2023swinia, li2021self}. Therefore, this paper divides the self-supervised image denoising algorithm into: General methods, BSN-based methods, Transformer-based methods.

\begin{table*}[t]
	\setlength{\abovecaptionskip}{0cm}
	\setlength{\belowcaptionskip}{0cm}
	\caption{General self-supervised image denoising method.}
	\centering
	\begin{tabular}{|m{2.2cm}<{\raggedright}|m{2cm}<{\raggedright}|m{4cm}<{\raggedright}|m{7.5cm}<{\raggedright}|}
		\hline
		Method&	Other needs	&Applications (denoising type)	&Key words (remarks)\\
		\hline
		N2N\cite{lehtinen2018noise2noise}&	Paired noise images&	Gaussian, Poisson, Bernoulli noise denoising and random text overlays remove.&	Real strictly aligned noiay-noisy image pairs.\\
		\hline
		GCBD \cite{chen2018image}&	Unpaired clean images&	Real-world sRGB image noise, Gaussian and Mixture noise denoising.&	GAN learned to generate noise distribution from noisy image and adding the noise to the clean image to get noisy-clean pairs.\\
		\hline
		SURE-based Method\cite{soltanayev2018training}&	Noise Model&	Gaussian noise denoising.&	Stein’s unbiased risk estimator (SURE), a SURE-based refining method.\\
		\hline
		Noisier2noise \cite{moran2020noisier2noise}&	Arbitrary noise model&	Gaussian additive noise and multiplicative Bernoulli noise denoising.&	Add synthetic noise signals to the given noise image as labels and continue add same type of noise to the labels as inputs. \\
		\hline
		NAC \cite{xu2020noisy}	&Noise model & AWGN and real-world sRGB image noise denoising.&	Add synthetic noise signals to the given weak noise images as inputs, weak noise images as targets.\\
		\hline
		R2R \cite{pang2021recorrupted}&	Noisy level function(NLF) or ISP function&	AWGN and real-world sRGB image noise denoising.&	Use data augmentation technique to get noisy-noisy pairs.\\
		\hline
		NBR2NBR \cite{huang2021neighbor2neighbor}&	$\backslash$	&Gaussian, Poisson  noise and real-world rawRGB image noise denoising.&	Two sub noise images that get from one noise image can be regard as a noisy-noisy pair.\\
		\hline
		Noise2Score \cite{kim2021noise2score}&	Arbitrary noise model&	Gaussian, Poisson, and Gamma noise denoising.&	The first step is to train a Neural Network (NN) to estimate the score function and then obtain the final denoising result by Tweedie’s formula. It can be used to deal with any exponential family noises.\\
		\hline
		Kim et al. \cite{kim2022noise}&	Arbitrary noise model&	Gaussian, Poisson, Gamma noise denoising.&	First to learn the score matching by a NN; second to obtain the denoised result via distribution-independent Tweedie’s formula, a new noise model and noise parameters estimation algorithm.\\
		\hline
		CVF-SID \cite{neshatavar2022cvf}&	$\backslash$ &	Real-world sRGB image noise denoising.&	Cyclic multi-Variate Function (CVF), uses the designed CNN model decompose the sRGB noise image to clean image, signal-independent and signal-dependent noise images.\\
		\hline
		IDR \cite{zhang2022idr}&	Noise model	&Gaussian, binomial and impulse noise, real-world raw noise denoising.&	Improving model denoising performance with an iterative method.\\
		\hline
		Vaksman et al. \cite{vaksman2023patch}&	$\backslash$ &	Gaussian noise, real-world sRGB image noise denoising.&	Capture bursts of noisy shots, and one is the input, and the rest are conducted by patch matching and stitching as targets, forming one-multiple training pairs.\\
		\hline
	\end{tabular}
	\label{tab:general}
\vspace{-2em}
\end{table*}

\subsection{General self-supervised image denoising methods}
\noindent Lehtinen et al. \cite{lehtinen2018noise2noise} proposed a self-supervised image denoising method called Noise2Noise (N2N)\cite{lehtinen2018noise2noise}, based on the UNet \cite{ko2018pac} structure. N2N requires perfectly aligned noisy-noisy image pairs with the same image signal $c$ but different noises $n_1$ and $n_2$. Paired noise images $x_1$= $c+n_1$, $x_2= c+n_2$ are used as input and target for training respectively, and the loss function can be expressed as:
\begin{equation}
	\setlength{\abovedisplayskip}{0pt}
	\setlength{\belowdisplayskip}{0pt}
	L(\theta) = \Vert F(c+n_1;\theta) - (c+n_2) \Vert^2
	\label{eq:N2N}
\end{equation}
where $F(\cdot)$ denotes the UNet model and $\theta$ denotes the model parameters. Through such a strategy, N2N achieves a denoising effect comparable to supervised image denoising. However, N2N requires noisy-noisy image pairs that are strictly aligned which can be difficult to obtain in practice due to environmental factors such as wind, sunlight, etc. To overcome this, scholars use noise models \cite{moran2020noisier2noise, krull2020probabilistic, xu2020noisy, kim2021noise2score, zhang2022idr, ramani2008monte} or noise level estimation \cite{chen2018image, zhang2022idr, kim2022noise} to artificially generate noisy-noisy pairs. Some use  unpaired noisy-clean images to obtain simulated noise \cite{chen2018image, wu2020unpaired, moran2020noisier2noise} and add simulated noise to the clean images for denoising model training, while others only require single noisy images for denoising model training\cite{krull2019noise2void,batson2019noise2self,laine2019high, xie2020noise2same, quan2020self2self, huang2021neighbor2neighbor, wang2022blind2unblind,lee2022ap,neshatavar2022cvf, zhang2023mm, vaksman2023patch, li2023spatially}.

\subsubsection{Unpaired noisy-clean images methods}
\
\newline
\noindent GCBD \cite{chen2018image} is an image denoising method that uses unpaired noisy-clean images, and is divided into two stages: noise level estimation and noise removal. First, noise blocks are obtained from noise images, and a generative adversarial network (GAN) is trained to estimate the noise distribution and generate noise samples. Second, the generated noise samples are added to the clean image to obtain a synthetic noisy image to construct a paired training dataset, which is used to train a CNN to denoise a given noisy image.
To obtain a set of approximate noise blocks $V$, GCBD subtracts the average of relatively smooth blocks in the noisy image under the assumption that the noise distribution has zero-mean. 
GCBD uses a WGAN-GP \cite{gulrajani2017improved} loss for the GAN and a CNN with a structure similar to DnCNN \cite{zhang2017beyond}. The loss for the GAN and CNN are described as following Eq. (\ref{eq:GCBD_1}, \ref{eq:GCBD_2}):
\begin{equation}
	\setlength{\abovedisplayskip}{0pt}
	\setlength{\belowdisplayskip}{0pt}
	\begin{aligned}
		L_{GAN} &= E_{\tilde{x}\sim P_g}[D(\tilde{x})]-E_{x\sim P_r}[D(x)] \\
		&+ \gamma E_{\hat{x}\sim P_{\hat{x}}}[(\Vert \nabla_{\hat{x}} D(\hat{x})\Vert_2 -1)^2]
	\end{aligned}
	\label{eq:GCBD_1}
\end{equation}
where $P_r$ is the distribution of $V$, $P_g$ is the generator distribution, $P_{\hat{x}}$ is defined as the distribution uniformly sampled along the line between pairs of points sampled from $P_r$ and $P_g$ \cite{arjovsky2017wasserstein}, and $D(\cdot)$ denotes the output of the discriminative network regarding the corresponding noise block as input. 
\begin{equation}
	\setlength{\abovedisplayskip}{0pt}
	\setlength{\belowdisplayskip}{0pt}
	L_{CNN}(\theta) = \frac{1}{2N} \sum\limits_{i=1}^N\Vert R(y_i;\theta) - (y_i-x_i) \Vert_{F}^{2}
	\label{eq:GCBD_2}
\end{equation}
where $R(\cdot)$ denotes the output of the CNN, $x_i$ is the clean image, $y_i$ is the synthetic noisy image, $N$ is the batch size of the training and $\theta$ denotes the network parameters.
GCBD does not require paired noisy-clean or noisy-noisy images, but only unpaired noisy-clean images for training. The noise synthesized by the GAN is closer to real noise than previous supervised methods that use specific noise models. However, GCBD assumes additive noise with zero-mean, which may not cover all types of noise, such as multiplicative noise.

Soltanayev et. al \cite{soltanayev2018training} proposed a self-supervised image denoising method based on Therorem of Monte-Carlo Stein’s unbiased risk estimator (MC-SURE \cite{ramani2008monte}) as Eq. (\ref{eq:SURE_1}, \ref{eq:SURE_2}):
\begin{equation}
	\setlength{\abovedisplayskip}{0pt}
	\setlength{\belowdisplayskip}{0pt}
	\sum\limits_{i=1}^K\frac{\partial h_i(y)}{\partial y_i}= \lim \limits_{\epsilon \rightarrow 0}{E_{\tilde{n}}\{\tilde{n}^t(\frac{h(y+\epsilon \tilde{n})-h(y)}{\epsilon})\}}
	\label{eq:SURE_1}
\end{equation}
where $\tilde{n} \sim \aleph_{0,1} \in \mathbb{R}^K$ be independent of $n$ and $y$, and provided that $h(y)$ admits a well-defined sencond-order Taylor expansion. If not, this is still valid in the weak sense provided that $h(y)$ is tempered.
So,
\begin{equation}
	\setlength{\abovedisplayskip}{0pt}
	\setlength{\belowdisplayskip}{0pt}
	\frac{1}{K}\sum\limits_{i=1}^K\frac{\partial h_i(y)}{\partial y_i} \approx \frac{1}{\epsilon K}\tilde{n}^t(h(y+\epsilon \tilde{n})-h(y))
	\label{eq:SURE_2}
\end{equation}
where $\epsilon$ is a fixed small positive value, $t$ is the transpose operator.
The SURE-based method proposes to incorporate MC-SURE \cite{ramani2008monte} into a stochastic gradient-based optimization algorithm for image denoising. An unbiased estimator is proposed to replace the unknown Bayesian risk for training, and with a carefully defined cost function, a learning-based image denoiser can be trained using this SURE-based method without requiring a noiseless ground truth. In addition, other deep learning-based denoisers can also utilize the MC-SURE-based training by modifying their cost function to a stochastic gradient-based optimizing format, as long as it satisfies the MC-SURE.
This method assumes Gaussian noise with known variance in all simulations, but it can incorporate a variety of noise distributions like Poisson distribution and exponential families, potentially making it applicable to different applications in the measurement domain. However, the method is sensitive to hyper-parameters, and the neural network needs to be retrained if the underlying noise model changes.

C2N \cite{jang2021c2n} framework trains a noise generator network $G$ to create a realistic noise map $\hat{n}$ for a given clean image $x$, which produces pseudo-noise images $\hat{y}$ as follows:
\begin{equation}
	\setlength{\abovedisplayskip}{0pt}
	\setlength{\belowdisplayskip}{0pt}
	\hat{y} = x+ \hat{n} = x+ G(x,r)
	\label{eq:C2N_1}
\end{equation}
The 32-dim random vector $r$ is sampled from $\mathcal{N}(\bf{0}, \bf{I^2})$ to reflect the stochastic behavior of noise according to the conditions of each scene, and authors use a new generator architecture to extract the diverse and complex noise  distributions combining the Signal-Independent Pixel-Wise Transform, Signal-Dependent Pixel-Wise Transform and Spatially Correlated Transforms.
A discriminator network $D$ distinguishes whether a given noisy image is synthesized from the generator $G$ or sampled from the real-world dataset. The two networks $G$ and $D$ are optimized adversarially with the Wasserstein distance \cite{gulrajani2017improved} to learn the noise distribution as follows:
\begin{equation}
	\setlength{\abovedisplayskip}{0pt}
	\setlength{\belowdisplayskip}{0pt}
	\begin{aligned}
		L_{adv}(D,G) &=E_{y^{\prime} \sim P_N}[D(y^{\prime})] \\
		&-E_{x\sim P_C, r \sim P_r}[1-D(x+G(x,r))] \\
		&+ \lambda E_{x_{\delta}\sim P_{\delta}}[(\Vert \nabla_{x_{\delta}} D(x_{\delta})\Vert_2 -1)^2]
	\end{aligned}
	\label{eq:C2N_2}
\end{equation}
$P_N$ and $P_C$ denote the distribution of the real-world noisy and clean images, respectively. The real noisy image $y^{\prime}$ is sampled from $P_N$. Prime notation on $y$ denotes that we use noisy image unpaired to clean image $x$. $P_r$ is the distribution of random vector $r$.
The authors use a stabilizing loss term $L_{stb}$ to prevent the color-shifting problem, which is defined as follows:
\begin{equation}
	\setlength{\abovedisplayskip}{0pt}
	\setlength{\belowdisplayskip}{0pt}
	L_{std} = \frac{1}{N}\sum \limits_{c} \Vert \sum\limits_{i \in B}{\hat{n}_{i,c} \Vert_1}
	\label{eq:C2N_3}
\end{equation}
where $N$ denotes the number of pixel $i$ in mini-batch $B$ and $c$ is index of each color channels. They minimize this loss to make the channel-wise average of the generated noise approach zero. 
For denoising, the C2N framework uses pseudo-noise images $\hat{y}$ to train a denoising model in a supervised manner with $L_1$ loss. However, this denoising method may not perfectly reflect real-world noise, leading to potential deviations in practical applications.

\subsubsection{Methods relying on artificially generated noisy-noisy pairs}
\
\newline
\noindent Noisier2Noise \cite{moran2020noisier2noise} was introduced in 2020 that was a variant of the N2N \cite{lehtinen2018noise2noise} method. It uses two fully aligned noise images for denoising model training, but these images are not collected naturally. Instead, the method adds the same noise type but different noises successively to a noise image.
This process generates an image pair $(Z,Y)$, with $Z$ as the input for the network $f(\cdot)$ and $Y$ as the training target. Then:
\begin{equation}
	\setlength{\abovedisplayskip}{0pt}
	\setlength{\belowdisplayskip}{0pt}
	Y \triangleq  X + N
	\label{eq:Noisier2Noise_1}
\end{equation}
\begin{equation}
	\setlength{\abovedisplayskip}{0pt}
	\setlength{\belowdisplayskip}{0pt}
	Z \triangleq  Y+M = X+N +M
	\label{eq:Noisier2Noise_2}
\end{equation}
The loss function used for training is $L_2$ loss:
\begin{equation}
	\setlength{\abovedisplayskip}{0pt}
	\setlength{\belowdisplayskip}{0pt}
	L(\theta) = \min\limits_{\theta}{E_Z \Vert f(Z;\theta) -Y\Vert_2}
	\label{eq:Noisier2Noise_3}
\end{equation}
Since $M$ and $N$ belong to the same known noise model, they are approximately equal, it can be deduced that:
\begin{equation}
	\setlength{\abovedisplayskip}{0pt}
	\setlength{\belowdisplayskip}{0pt}
	E[X|Z] = 2E[Y|Z] -Z
	\label{eq:Noisier2Noise_4}
\end{equation}
After training, the original unknown noise $n$ can be obtained by subtracting $Z$ from twice the model output. The denoised image can be obtained by subtracting $n$ from the original noise image.
Noisier2Noise only needs single noise images and a noise distribution statistical model to train the denoising model. It can also denoise spatially correlated noise. However, it requires a known noise model, which may not always be available or accurate in practice. Additionally, the noise model used for training may not be representative of the noise in the actual images, leading to suboptimal denoising results. Finally, Noisier2Noise may not perform well on images with complex noise patterns.

Noisy-As-Clean (NAC) \cite{xu2020noisy} assumes the noise in the image is weak enough, so the noise image can be treated as a clean image for denoising model training. NAC uses paired noise images $(y, z)$ generated by the clean image $x$ and synthetic noise $n_s$ that has the same type of noise as the natural weak noise $n_0$. The input
$y$ and the target $z$ can be described as:
\begin{equation}
	\setlength{\abovedisplayskip}{0pt}
	\setlength{\belowdisplayskip}{0pt}
	y = x+ n_0
	\label{eq:NAC_1}
\end{equation}
\begin{equation}
	\setlength{\abovedisplayskip}{0pt}
	\setlength{\belowdisplayskip}{0pt}
	z=y+n_s
	\label{eq:NAC_2}
\end{equation}
\begin{equation}
	\setlength{\abovedisplayskip}{0pt}
	\setlength{\belowdisplayskip}{0pt}
	E[x] >> E[n_0], Var[x]>> Var[n_0]
	\label{eq:NAC_3}
\end{equation}
$E(\cdot)$ denotes the expectation and $Var(\cdot)$ denotes the variance. Hence, the noisy image $y$ should have similar expectation with the clean image $x$,
\begin{equation}
	\setlength{\abovedisplayskip}{0pt}
	\setlength{\belowdisplayskip}{0pt}
	E[x] >> E[n_s], Var[x]>> Var[n_s]
	\label{eq:NAC_4}
\end{equation}
then,
\begin{equation}
	\setlength{\abovedisplayskip}{0pt}
	\setlength{\belowdisplayskip}{0pt}
	E[y]=E[x + n_0]=E[x]+E[n_0] \approx E[x]
	\label{eq:NAC_5}
\end{equation}
\begin{equation}
	\setlength{\abovedisplayskip}{0pt}
	\setlength{\belowdisplayskip}{0pt}
	E[z] = E[y+n_s] \approx E[y]
	\label{eq:NAC_6}
\end{equation}
so,
\begin{equation}
	\setlength{\abovedisplayskip}{0pt}
	\setlength{\belowdisplayskip}{0pt}
	E_y[E_z[z|y]] = E[z] \approx E[y] = E_x[E_y[y|x]]
	\label{eq:NAC_7}
\end{equation}
As can be seen from the above, in theory, when the noise in the image is weak enough, it can be treated as a clean image for denoising model training as the target.
NAC can be combined with any CNN model, and it can achieve better performance than methods that use noisy-clean image pairs, especially when the noise is weak. However, NAC requires prior knowledge of the noise type and noise level estimation, which can be difficult to obtain in the real world. Although the article mentions using the Poisson-Gaussian mixture distribution when the noise type is unknown, many noises do not follow this distribution in practice. 

Recorrupted-to-Recorrupted (R2R) \cite{pang2021recorrupted} was introduced in 2021 that used a data augmentation method to train a denoising model with only single noisy images. Like N2N \cite{lehtinen2018noise2noise} that uses noisy-noisy image pairs to train the model, R2R generates these images from a single noisy image using different data augmentation methods for independent noise and noise correlated with the image signal, respectively.
For independent noise, paired noise images $(\hat{y}, \tilde{y})$ can be generated by the following:
\begin{equation}
	\setlength{\abovedisplayskip}{0pt}
	\setlength{\belowdisplayskip}{0pt}
	\hat{y}=y+D^Tz, \tilde{y} = y-D^{-1}z, z \sim \mathcal{N}(\bf{0},\bf{\sigma^2I})
	\label{eq:R2R_1}
\end{equation}
where $y$ is the noisy image, $D$ can be any invertible matrix and $z$ is sampled from standard normal distribution $\mathcal{N}$.
For noise correlated with the image signal $x$, paired noise images can be generated using a similar formula with the $x$-dependent covariance matrix $\sqrt{\sum_x}$ by the Eq. (\ref{eq:R2R_2}):
\begin{equation}
	\setlength{\abovedisplayskip}{0pt}
	\setlength{\belowdisplayskip}{0pt}
	\hat{y}=y+\sqrt{\sum_x}D^Tz, \tilde{y} = y-\sqrt{\sum_x}D^{-1}z, z \sim \mathcal{N}(\bf{0,I})
	\label{eq:R2R_2}
\end{equation}
R2R achieved competitive results for both AWGN removal and real-world image denoising compared to the state-of-the-art self-supervised learning methods. However, it requires a significant amount of computational resources due to the data augmentation method used for training, and R2R requires prior knowledge of the noise level, which can be difficult to obtain in the real world.

Zhang et al. \cite{zhang2022idr} proposed the Iterative Data Refinement (IDR) method trained with noiser-noise image pairs. First, IDR proves that the lower the noise level on the target noisy image the better the denoising effect. Therefore, it is feasible to reduce the noise level on the original noisy image to improve the denoising performance of the model. Then, IDR is proposed to be divided into two steps: (1) train a model $M_0$ with noise reduction ability through the noisier-noise pairs $\{y_0+n$, $y_0\}$. (2) iteratively replace the noisy image $y_0$ with the preliminary denoising image from the previous round $M_0$ until the loss function converges.
Through iteration in this way, the denoising performance of the model is continuously improved. However, this iterative method is easy to lose the detail information in the image signal, resulting in the final denoised image being too smooth.

Noise2Score \cite{kim2021noise2score} is a denoising method that estimates the score function of an image using Bayesian statistics, specifically Tweedie's formula \cite{efron2011tweedie}, which can be applied to any exponential family of noise. The method involves adding certain noise at different levels to noisy images to estimate the score function, rather than the noise itself. A post-processing step with Tweedie's formula is then used, which is determined by the specific noise model, such as Poisson or Gamma, etc.
Noise2Score is more flexible than some other denoising methods that assume a specific type of noise. Additionally, it estimates the score function directly rather than the noise itself, and it can achieve better denoising performance than methods that rely on estimating the noise level.

Kim et al. \cite{kim2022noise} developed a denoising method that uses a quadratic equation for noise model estimation, which can be applied to Gaussian, Poisson, and Gamma noise models. The method requires only one additional inference step for noise level estimation, which is more efficient than Noise2Score \cite{kim2021noise2score}, which requires multiple inferences. 
The method involves adding unknown noise to noisy images and estimating the noise model using the quadratic equation. The estimated noise model is then used to estimate the noise level and denoise the image.
However, it assumes that the unknown noise is limited to these three common noise models, so it may not perform well with other types of noise.

\subsubsection{Methods relying on only single noisy images}
\
\newline
\noindent Neighbor2Neighbor (NBR2NBR) \cite{huang2021neighbor2neighbor} is introduced in 2021 that uses a specific down-sampling strategy to divide the entire noise image into multiple $2\times2$ cells, achieving noisy-noisy image pairs. The method randomly selects two pixels in each cell to get two sub-images and uses them as the input and target respectively for denoising model training. To prevent overfitting, a regularization term is added to supplement lost details.
The loss function for NBR2NBR includes a reconstruction term and a regularization term as formulas in Eq. (\ref{eq:NBR2NBR_1}-\ref{eq:NBR2NBR_3}), which help to preserve image details:
\begin{equation}
	\setlength{\abovedisplayskip}{0pt}
	\setlength{\belowdisplayskip}{0pt}
	L_{recon} = \Vert F(g_1(y)) -g_2(y)\Vert_{2}^{2}
	\label{eq:NBR2NBR_1}
\end{equation}
\begin{equation}
	\setlength{\abovedisplayskip}{0pt}
	\setlength{\belowdisplayskip}{0pt}
	L_{regular} = \Vert F(g_1(y)) - g_2(y) - F(g_1(F(y))) - g_2(F(y))\Vert_{2}^{2}
	\label{eq:NBR2NBR_2}
\end{equation}
\begin{equation}
	\setlength{\abovedisplayskip}{0pt}
	\setlength{\belowdisplayskip}{0pt}
	L = L_{recon} + \gamma L_{regular}
	\label{eq:NBR2NBR_3}
\end{equation}
where $F(\cdot)$ represents the denoising model, $g_1 (y)$ and $g_2 (y)$ respectively represent the two sub-images obtained by applying the mask strategy on the noise image $y$, $g_1 (F(y))$ and $g_2 (F(y))$ respectively represent two sub-images generated by using the down-sampling strategy on the output of the denoising model when the noise image $y$ is used as input with none gradient.
Synthetic experiments with different noise distributions in sRGB space and real-world experiments on a denoising benchmark dataset in raw-RGB space have shown that NBR2NBR can achieve state-of-the-art denoising performance.
However, approximating neighbor pixels can lead to over-smoothing, and down-sampling can destroy structural continuity.

CVF-SID\cite{neshatavar2022cvf} was introduced in 2022 by Reyhaneh Neshatavar et al. that proposes a Cyclic multi-Variate Function (CVF) module. The method only requires single noisy images. CVF-SID first decomposes the noisy imag  $I_n$ into a clean image $I_c$, signal-dependent noise $N_d$ and signal-independent noise $N_i$: 
\begin{equation}
	\setlength{\abovedisplayskip}{0pt}
	\setlength{\belowdisplayskip}{0pt}
	I_c,I_d,N_i = F(I_n)
	\label{eq:CVF_1}
\end{equation}
where $F(\cdot)$ denotes a CNN model. And then it combines the three output parts arbitrarily as a new input to the previous model for decomposition training,
\begin{equation}
	\setlength{\abovedisplayskip}{0pt}
	\setlength{\belowdisplayskip}{0pt}
	I_{n_2} = g(s_1I_c, s_2N_d, s_3N_i)
	\label{eq:CVF_2}
\end{equation}
\begin{equation}
	\setlength{\abovedisplayskip}{0pt}
	\setlength{\belowdisplayskip}{0pt}
	I_{c_2}, N_{d_2}, N_{i_2}=F(I_{n_2})
	\label{eq:CVF_3}
\end{equation}
where $g$ is a combination function, $s_1$, $s_2$, $s_3$  are constants which can be one value from [-1, 0, 1]. Finally, repeat this cycle until the loss converges.
To allow the CNN model to decompose noisy images, the method uses various self-supervised loss terms based on statistical behaviors of general noise. 
CVF-SID directly decomposes the sRGB noise image, maintaining the texture structure of the image better than BSN-based methods.
However, the method relies on the assumption that the noise in the image can be decomposed into signal-dependent and signal-independent components, which may not always be true in practice. Therefore, the method may not perform well on images with noise that is not well-modeled by this assumption. Additionally, the cyclic training process used by CVF-SID is computationally expensive and may take longer to converge than other denoising methods. This may make it difficult to use in real-time applications or on large datasets. Finally, while CVF-SID is able to maintain the texture structure of the image better than some other methods \cite{krull2019noise2void, batson2019noise2self, pang2021recorrupted}, it may still introduce some degree of smoothing or loss of detail due to the decomposition and recombination process.

Vaksman et al. \cite{vaksman2023patch} presented the Patch-Craft Self-Supervised Training for Correlated Image Denoising paper at CVPR 2023, to address the challenge of denoising images with an unknown noise model. The method uses bursts of noisy images for self-supervised denoiser training, where one image from the burst is used as input and the rest are used for constructing targets. The targets are constructed using the patch-craft frames concept introduced in PaCNet \cite{efron2011tweedie}, where the input shot is split into fully overlapping patches and the nearest neighbor within the rest of the burst images is found for each patch. It is important to note that the input shot is strictly excluded from the neighbor search.
The paper also proposes a novel method for performing the statistical analysis of the target noise, which involves cutting the proper left tail shown on the histogram of an empirical covariance for the input and the difference between target and input. This method allows trainers to exclude faulty image pairs from the training set, which can boost the performance of the denoising method.
One potential limitation of the method is that it requires bursts of noisy images for training, which may not always be available in real-world scenarios. Additionally, the method may not perform well on images with highly complex noise patterns that cannot be effectively modelled by the patch-craft frames concept. Furthermore, the method may be computationally expensive due to the neighbor search process used for constructing targets.

\subsection{BSN-based self-supervised image denoising methods}
\noindent BSN is a method of denoising images by predicting the noisy-free pixels of the masked pixels using the spatial continuity between the masked pixels and their surrounding pixels in the image signal. The effectiveness of BSN is based on the assumption that the noise in the image is spatially independent and zero-mean, while the image signal exhibits spatial correlation. 

\begin{figure*}[htbp]
	\setlength{\abovecaptionskip}{0cm}
	\setlength{\belowcaptionskip}{0cm}
	\centering
     \begin{minipage}[b]{0.28\linewidth}
		\subfloat[N2V\cite{krull2019noise2void}, P2V\cite{krull2020probabilistic}]{\label{fig:figure 2a}
		\includegraphics[height=2.0in]{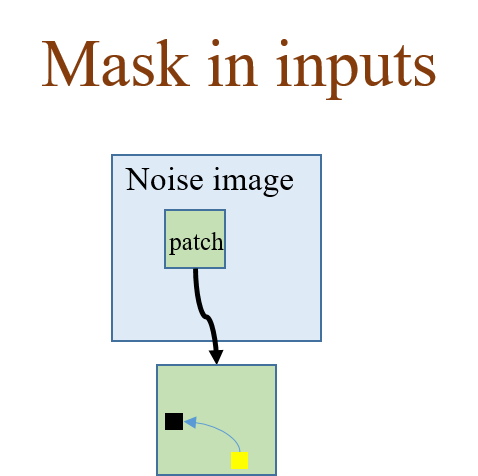}}
	\end{minipage}
	\begin{minipage}[b]{0.65\linewidth}
		\subfloat[N2S\cite{batson2019noise2self}, Noise2Same\cite{xie2020noise2same}, S2S\cite{quan2020self2self}]{\label{fig:figure 2b}
		\includegraphics[height=0.85in]{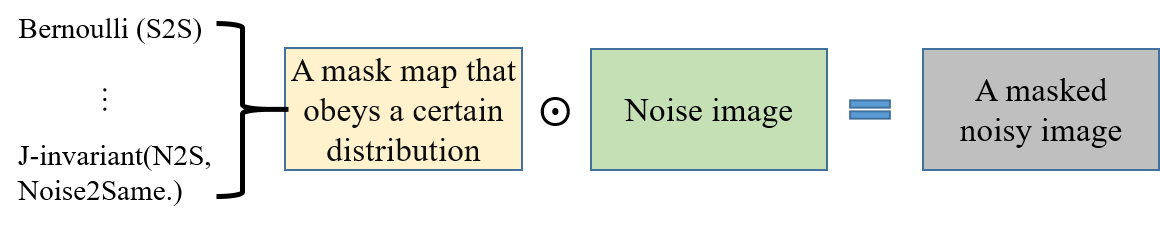}} 
           \quad
		\subfloat[B2UB\cite{wang2022blind2unblind}, DT\cite{zhang2023self}]{\label{fig:figure 2c}
		\includegraphics[height=0.85in]{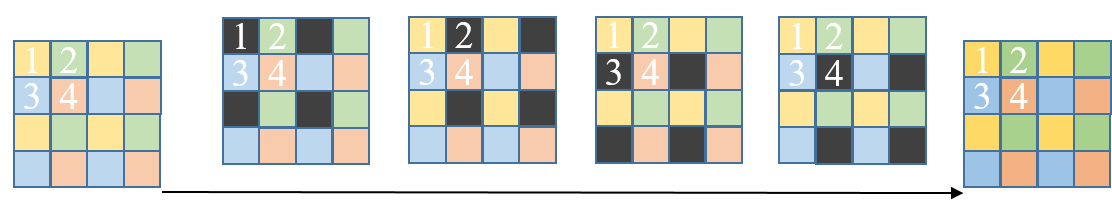}} 
	\end{minipage} 
	\caption{Different mask in input strategies}
	\label{fig:figure 2}
\vspace{-1em}
\end{figure*}

BSN-based methods can be divided into two categories based on the mask strategy: mask in input \cite{krull2019noise2void, batson2019noise2self, krull2020probabilistic, xie2020noise2same, quan2020self2self, kim2021noise2score, kim2022noise, wang2022blind2unblind, zhang2023self, ramani2008monte} and mask in network \cite{laine2019high, wu2020unpaired, lee2022ap, wang2023lg, zhang2023mm, li2023spatially}. Mask in input uses a specific strategy to mask some pixels on the noisy image as input, and the complete noisy image as the target to perform supervised training on deep neural networks. In contrast, mask in network involves masking part of the receptive field, including the center pixel, during feature extraction in the network structure, so that the model uses some surrounding pixels of the feature maps to predict the target pixel. These two techniques are illustrated in the Fig. \ref{fig:figure 2}-\ref{fig:figure 3}, respectively.

\begin{table*}[t]
	\setlength{\abovecaptionskip}{0cm}
	\caption{BSN-based self-supervised image denoising methods of mask in inputs.}
	\centering
	\begin{tabular}{|m{2.2cm}<{\raggedright}|m{2cm}<{\raggedright}|m{4cm}<{\raggedright}|m{7.5cm}<{\raggedright}|}
		\hline
		Method&	Other needs	&Applications (denoising type)	&Key words (remarks)\\
		\hline
		N2V\cite{krull2019noise2void}&	$\backslash$	&Gaussian noise and some biomedical image noise denoising.	&Pixel-wise independent noise, randomly select several pixel to mask in the input images.\\
		\hline
		N2S\cite{batson2019noise2self}&	$\backslash$	&Blind Gaussian noise denoising.&	 $\mathcal{J}$-invariant function determines the mask distribution, and replaces the pixels at  $\mathcal{J}$ with random numbers.\\
		\hline
		PN2V \cite{krull2020probabilistic}&	An arbitrary noise model&	Microscopy and low-light condition image noise denoising.&	Mask input images, probabilistic model, predict per-pixel intensity distributions.\\
		\hline
		Noise2Same \cite{xie2020noise2same}&	$\backslash$	&Gaussian noise denoising.	& $\mathcal{J}$-invariant function determines the mask distribution, and replaces the pixels at  $\mathcal{J}$ with local averages.\\
		\hline
		S2S \cite{quan2020self2self}&	$\backslash$	&Blind Gaussian, salt-and-pepper and real-world sRGB image noise denoising.&	Bernoulli-sampled instances of the input image results on noisy pairs.\\
		\hline
		B2UB\cite{wang2022blind2unblind}&	$\backslash$	&FMDD, Gaussian, Poisson and real-world rawRGB image noise denoising.&	Gobal-aware mask mapper, re-visible loss.\\
		\hline
	\end{tabular}
	\setlength{\belowcaptionskip}{-0.5cm}
	\label{tab:mask_in_input}
\vspace{-2em}
\end{table*}
\subsubsection{Mask in input methods}
\
\newline
\noindent The first BSN-based method, Noise2Void (N2V), was proposed by Krull et al. \cite{krull2019noise2void} in 2018. N2V is a self-supervised image denoising algorithm that only requires a single noisy image to train the model. During training, N2V randomly selects multiple pixels on each $64\times64$ patch in the noisy image to be replaced by any adjacent pixels, as shown in Fig. \ref{fig:figure 2a}. The resulting image with multiple pixels masked is regarded as input, and unmasked noisy image is regarded as the target for training. The corresponding loss function for N2V can be described as:
\begin{equation}
	\setlength{\abovedisplayskip}{0pt}
	\setlength{\belowdisplayskip}{0pt}
	L(\theta) = \Vert F(y_{masked};\theta) - y_{original} \Vert^2
	\label{eq:N2V}
\end{equation}
where $y_{original}$ represents the noisy image, $y_{masked}$ represents the masked version of $y_{original}$, and $\theta$ represents the parameters of the denoising model $F(\cdot)$. By randomly replacing some pixels in $y_{original}$ with adjacent pixels to create $y_{masked}$, the model is trained to predict the missing pixels based on their spatial continuity with the surrounding pixels. This approach prevents the model from simply memorizing the noisy image and encourages it to learn the underlying structure of the image. 
While N2V can obtain a denoising model with only a single noisy image, it has some limitations. Only the masked pixels participate in model training each time, which wastes training resources and time. Additionally, the masked pixels no longer appear in the training process, resulting in the loss of information during model training.

Noise2Self (N2S) \cite{batson2019noise2self} is proposed based on the assumption of noise with statistical independence.
The authors first demonstrate that the optimal denoising function for a given dataset can be found by minimizing the self-supervised loss using a class of $\mathcal{J}$-invariant functions, which are functions that preserve the structure of the input image under arbitrary translations. They then show through examples that as the degree of correlation between features in the data increases, the optimal $\mathcal{J}$-invariant denoising function approaches the optimal general denoising function.
Based on this observation, the authors propose a deep neural network with millions of parameters that is modified to become $\mathcal{J}$-invariant through the masking of pixels with random values. This involves adding a $\mathcal{J}$-invariant mask $\mathcal{J}$ to the input, as shown in Fig. \ref{fig:figure 2b}, where the original values on $\mathcal{J}$ are masked and replaced by random values. The authors conduct experiments using two neural networks, UNet \cite{ko2018pac} and DnCNN \cite{zhang2017beyond} on three datasets including ImageNet \cite{deng2009imagenet} and CellNet \cite{cahan2014cellnet} with different noise mixture strategies. 
The experiments show that N2S achieves competitive performance with the same neural network architectures trained with clean targets (called Noise2Truth) and independently noisy targets (N2N \cite{lehtinen2018noise2noise}). However, many open questions remain regarding the optimal choice of $\mathcal{J}$ in practice. The construction of $\mathcal{J}$ must reflect the signal-dependent patterns and noise independence relations, which involves making a bias-variance trade-off in the loss of the relative sizes of each subset $J$ and increasing information for prediction.

Several denoising methods \cite{lehtinen2018noise2noise, krull2019noise2void, laine2019high}, including N2V \cite{krull2019noise2void}, assume that the noise present in an image is uniform, which is not always the case in reality. Probabilistic Noise2Void (PN2V) \cite{krull2020probabilistic} extends the original N2V method and addresses this limitation by incorporating a probabilistic model of the noise. PN2V models the noise distribution using a probabilistic approach, which allows it to handle non-uniform noise and capture the variability in the noise distribution across different regions of the image. PN2V builds on N2V by introducing a probabilistic noise model that incorporates the noise distribution as a prior into the loss function. This allows the model to learn the noise distribution from the data and produce more accurate denoising results. 
In the PN2V model training followed by N2V, noisy images are first randomly masked as inputs and original noisy images as targets, shown in Fig. \ref{fig:figure 2a}. The network is then trained to predict probability distribution of the pixel’s signal in the masked regions, considering only its surroundings, while also learning the noise model parameters from the training data. The loss function of PN2V can be expressed as:
\begin{equation}
	\setlength{\abovedisplayskip}{0pt}
	\setlength{\belowdisplayskip}{0pt}
	L = \mathop{\arg\min}\limits_\theta \sum\limits_{i=1}^n {- \ln(\frac{1}{K} \sum\limits_{k=1}^K) {p(x_i|s_{i}^{k})}}
	\label{eq:PN2V}
\end{equation}
$p(\cdot)$ is described by an arbitrary noise model, which in the case of PN2V \cite{krull2020probabilistic} is a histogram-based noise model. For each pixel $i$, the network directly predicts $K$ output values ($K=800$, $s_i^k$ denotes each value), which are interpreted as independent samples, and $x_i$ denotes the observed noisy input value. 
Empirical results have shown that PN2V \cite{krull2020probabilistic} outperforms N2V \cite{krull2019noise2void} in a variety of imaging applications, including fluorescence microscopy and electron microscopy. Furthermore, the probabilistic noise model allows PN2V to estimate the uncertainty of the denoised image, which is useful in downstream applications that require uncertainty quantification. 
However, in the real world, most noise sources do not strictly follow the Poisson-Gaussian distribution or histogram-based noise models used in PN2V \cite{krull2020probabilistic}. Therefore, the PN2V model trained with a specific noise distribution will have limited performance when denoising real-world noise.

Noise2Same\cite{xie2020noise2same} proved through experiments that the denoising functions trained through mask-based blind-spot methods are not strictly $\mathcal{J}$-invariant, meaning that they do not preserve the structure of the input image under arbitrary translations. This can lead to reduced denoising performance, particularly in cases where the noise is correlated or exhibits complex statistical properties. Furthermore, Noise2Same demonstrated that minimizing the Euclidean distance between the denoised image and the original image, expressed as $E_x \Vert f(x) -x \Vert^2$,  is not optimal for self-supervised image denoising.
Noise2same masked the noise image same as N2S \cite{batson2019noise2self} to a $\mathcal{J}$-invariant distribution, but replaced the masked pixel with local average values instead of random values. It proposed to take both the original noisy image and the $\mathcal{J}$-invariant masked image as inputs and output two residual noise. The network would be trained from learning the invariance loss between the two outputs.

Self2Self With Dropout (S2S) \cite{quan2020self2self} uses Bernoulli dropout to improve the performance of self-supervised denoising methods. S2S randomly masks instances of the noisy image using a Bernoulli process, which generates a set of image pairs for training the neural network. The Bernoulli dropout is applied using element-wise multiplication, and it is used in both the training and test phases to reduce prediction variance. 
To further improve performance, S2S uses partial convolution instead of standard convolution for re-normalization on sample pixels. Partial convolution is a convolutional operation that only convolves over the unmasked pixels, which helps to preserve the image structure and prevent artifacts.
The self-prediction loss in S2S is defined on the pairs of Bernoulli sampled instances of the input image. The loss function encourages the network to predict the unmasked pixels of the noisy image from the masked pixels, while also taking into account the noise distribution and the image structure. The Bernoulli dropout and partial convolution help to regularize the denoising process and improve the performance of the network.

Blind2Unblind (B2UB) \cite{wang2022blind2unblind} is based on blind spots proposed by Wang et al. in 2022. B2UB uses a global-aware mask mapper to ensure that all the information in the image is fully utilized and uses re-visible loss to help restore more details in the image.
B2UB’s mask strategy is called the global-aware mask mapper. Firstly, the noise image is divided into multiple $k\times k$ cells, and the pixels are marked with a number $i$ ($i \in [1,k^2]$, $k=4$, in the experiments) from left to right and top to bottom in each cell. By masking the points with the same number in the original image, $k\times k$ masked images with the same size as the original image are obtained as inputs, as shown in Fig. \ref{fig:figure 2c}. Secondly, the denoising model produces $k\times k$ outputs. The pixels at all blind spots in the output images are collected and put back in the previously recorded positions, resulting in an image of the same size as the original noisy image with each pixel denoised. This image is combined with the original noisy image for supervised training.
In addition to the supervised loss, B2UB uses a re-visible loss as a regularization loss item to make full use of the information in the noisy image for detail restoration. The whole loss can be described as:
\begin{equation}
	\setlength{\abovedisplayskip}{0pt}
	\setlength{\belowdisplayskip}{0pt}
	L_{blind} = \Vert M^{-1}(F(M(y)) - y \Vert_{2}^{2}
	\label{eq:B2UB_1}
\end{equation}
\begin{equation}
	\setlength{\abovedisplayskip}{0pt}
	\setlength{\belowdisplayskip}{0pt}
	L_{re-visible} = \Vert M^{-1}(F(M(y)) + \alpha \widetilde{F}(y) -(1 + \alpha)y \Vert_{2}^{2}
	\label{eq:B2UB_2}
\end{equation}
\begin{equation}
	\setlength{\abovedisplayskip}{0pt}
	\setlength{\belowdisplayskip}{0pt}
	L = L_{blind} + \beta L_{re-visible}
	\label{eq:B2UB_3}
\end{equation}
where $y$ is the original noise image, $F(\cdot)$ represents the denoising model, $\tilde{F}(\cdot)$ denotes the denoising model without gradient update during training, $M(\cdot)$ represents the first step of the global-aware mask mapper which creates blind spots, $M^{-1} (\cdot)$ represents the second step of the global-aware mask mapper which collects the denoised blind spots, $\beta$ is a fixed hyper-parameter (set as 1) that determines how much the blind term affects model denoising training, and $\alpha$ is a variable hyper-parameter (increase from 2 to 20 in training) that controls the strength of the visible part.
B2UB uses the mask strategy of global-aware mask mapper to blind all points in the noise image once, so that all pixels participate in denoising training and make full use of the information of all pixels in the image. The re-visible loss is used to allow the model to see the complete noise image and repair the destroyed texture information in the process of creating blind spots. Extensive experiments on sRGB images with synthetic noise and rawRGB images with real noise demonstrate the superior performance of B2UB. However, B2UB cannot handle real-world sRGB images with spatially correlated noise that does not satisfy the prerequisites of BSN-based denoising methods.

\begin{figure*}[htbp]
	\setlength{\abovecaptionskip}{0cm}
	\setlength{\belowcaptionskip}{0cm}
	\centering
     \begin{minipage}[b]{0.32\linewidth}
		\subfloat[Laine et al. \cite{laine2019high}]{\label{fig:figure 3a}
		\includegraphics[height=2.0in]{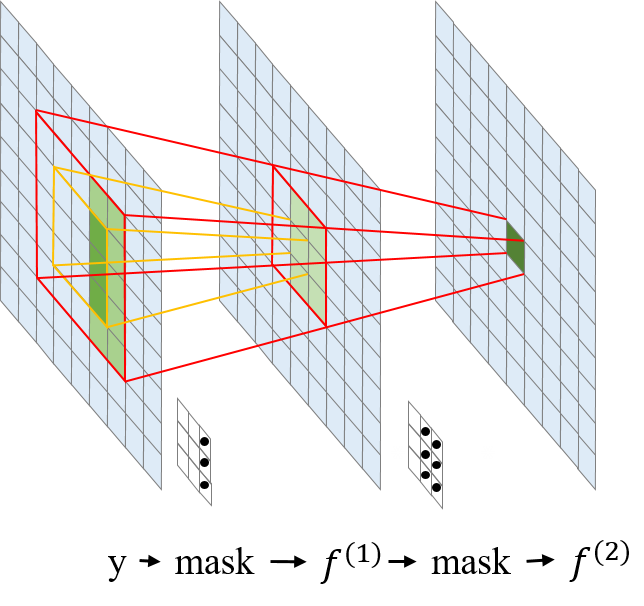}}
	\end{minipage}
     \begin{minipage}[b]{0.34\linewidth}
		\subfloat[DBSN\cite{wu2020unpaired}, AP-BSN\cite{lee2022ap}, LG-BSN\cite{wang2023lg}, Li et al.\cite{li2023spatially}]{\label{fig:figure 3b}
		\includegraphics[height=2.0in]{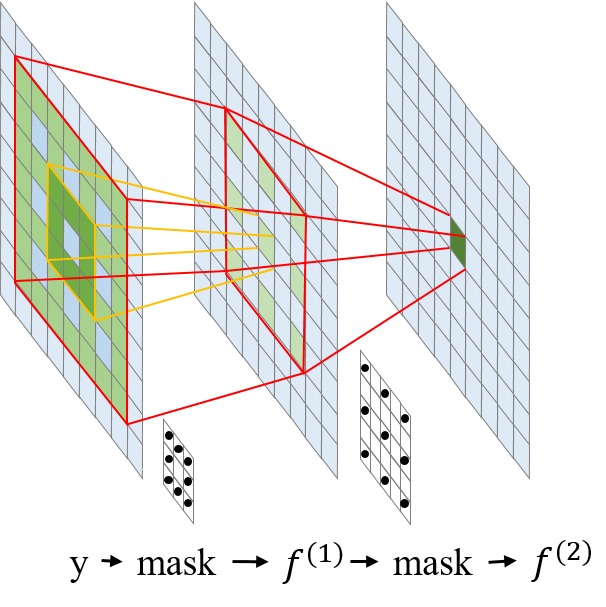}}
	\end{minipage}
     \begin{minipage}[b]{0.32\linewidth}
		\subfloat[MM-BSN\cite{zhang2023mm}]{\label{fig:figure 3c}
		\includegraphics[height=2.0in]{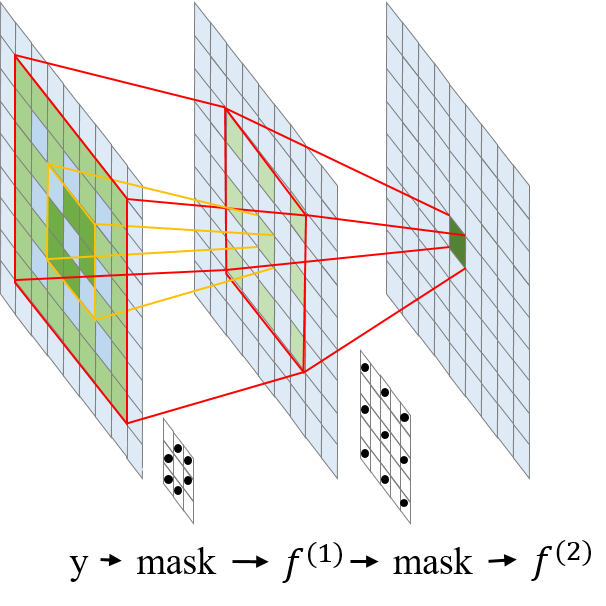}}
	\end{minipage}
	\caption{Different mask in network strategies}
	\label{fig:figure 3}
\vspace{-1em}
\end{figure*}

\begin{table*}[t]
	\setlength{\abovecaptionskip}{0cm}
	\setlength{\belowcaptionskip}{0cm}
	\caption{BSN-based self-supervised image denoising methods of mask in networks.}
	\centering
	\begin{tabular}{|m{2.2cm}<{\raggedright}|m{2cm}<{\raggedright}|m{4cm}<{\raggedright}|m{7.5cm}<{\raggedright}|}
		\hline
		Method&	Other needs	&Applications (denoising type)	&Key words (remarks)\\
		\hline
		Laine et al.\cite{laine2019high}&	$\backslash$	&Gaussian, Poisson and Impulse noise denoising.&	Mask four direction half receptive field respectively to achieve blind spot network.\\
		\hline
		DBSN \cite{wu2020unpaired}&	Unpaired clean images&	AWGN, heteroscedastic Gaussian (HG) and multivariate Gaussian (MG) noise and real-world sRGB image noise denoising.	&Dilated convolution, NLF, clean\underline{ }predict/noisy and clean/noisy\underline{ }synthetic pairs, knowledge distillation.\\
		\hline
		AP-BSN\cite{lee2022ap}&	$\backslash$	&Real-world sRGB image noise denoising.&	Asymmetric $PD$ in the training and testing, random-replacing refinement post-processing.\\
		\hline
		MM-BSN\cite{zhang2023mm}&	$\backslash$	&Real-world sRGB image noise denoising.&	Multi-mask strategy for large area spatial correlated noise, multi-mask combined network.\\
		\hline
		Li et al.\cite{li2023spatially}&	$\backslash$	&Real-world sRGB image noise denoising.&	Distinguish flat and textured regions in noisy images, and construct supervisions for them separately.\\
		\hline
	\end{tabular}
	\setlength{\belowcaptionskip}{-0.5cm}
	\label{tab:mask_in_networks}
\vspace{-2em}
\end{table*}

\subsubsection{Mask in network methods}
\
\newline
\noindent Laine et al. \cite{laine2019high} proposed a method for generating blind spots without using a mask on the input image for efficient training. The authors designed a novel architecture consisting of four parallel UNets\cite{ko2018pac} as the model framework. To achieve the effect of blind spots, four networks only look at the receptive fields of the upper, lower, left, and right half planes that do not include the central pixel, as shown in Fig. \ref{fig:figure 3a}.  Only a series of $1\times1$ convolutions are used to combine the output features of the four branches to ensure that the receptive field will not expand, thereby ensuring that the central pixel will never be seen in the entire network frame. The method of Laine et al. does not mask the input image, but the receptive field in the mask network framework, so that all pixels can participate in the training. Compared with N2V \cite{krull2019noise2void}, this method greatly improves the training efficiency and fully utilizes the information of all pixels. However, the large number of model parameters may limit the scalability of the method to larger images or more complex noise patterns.

In 2020, Wu et al. \cite{wu2020unpaired} proposed a two-stage image denoising method using unpaired clean $x$ and noisy $y$ images. They trained a denoising model with Dilated Blind-Spot Network (DBSN), as shown in Fig. \ref{fig:figure 3b} and knowledge distillation. 
In the first stage, noisy images are input into the proposed DBSN and the noise estimation model $CNN_{est}$ respectively. Through maximizing the constrained log-likelihood as the loss function to conduct self-supervised image denoising training, the denoised image and the noise level estimation function can be obtained. 
The loss function for the first stage when learning DBSN and $CNN_{est}$ is the constrained negative log-likelihood is:
\begin{equation}
	\setlength{\abovedisplayskip}{0pt}
	\setlength{\belowdisplayskip}{0pt}
	\begin{aligned}
		L_{self} &= \sum_{i} \{\frac{1}{2}(y_i-\hat{x_i})^T(\widehat{\sum}_{i}^{x} + \widehat{\sum}_{i}^{n})^{-1}(y_i -\hat{x_i}) \\
		&+ \log |\widehat{\sum}_{i}^{n}| + tr((\widehat{\sum}_{i}^{n})^{-1}\widehat{\sum}_{i}^{x}) \}
	\end{aligned}
	\label{eq:DBSN_1}
\end{equation}
where $tr(\cdot)$ denotes the trace of a matrix. $y_i$ is the real noisy image, $\hat{x}_i$ is the output of DBSN, $\widehat{\sum}_{i}^{n}$ and $\widehat{\sum}_{i}^{x}$ is a $C \times C$ covariance matrix output by $CNNest$ and D-BSN respectively for the position $i$. 
In the second stage, the DBSN denoising model trained in the first stage is used to obtain the denoised image $\hat{x}_i$ of the original noise image, and the noise level function is applied to the original clean image to obtain a synthetic noise image $\hat{y}_i$. Two pairs of image sets ($y_i$, $\hat{x}_i$) and ($\hat{y}_i$, $x_i$) are created, and they are input into any CNN model for joint training to obtain the final CNN denoiser. Its loss function is:
\begin{equation}
	\setlength{\abovedisplayskip}{0pt}
	\setlength{\belowdisplayskip}{0pt}
	\begin{aligned}
		L_{distill} &= \sum_{i} \Vert CDN(\hat{y_i}) - x_i \Vert^2 \\
		&+ \beta \sum_{i} \Vert CDN(y_i) -\hat{x_i} \Vert^2
	\end{aligned}
	\label{eq:DBSN_4}
\end{equation}
where $\beta = 0.1$ is the trade-off parameter, convolutional denoising network ($CDN$) can be any existing CNN denoisers.
Wu et al. \cite{wu2020unpaired} utilizes unpaired noisy-clean image pairs for training the denoising model. The use of real clean images in the training allows the model to focus on more detail restoration, and the real noisy images make the model more adaptable to the removal of noise in the real world. The two branches are jointly trained, and the final denoising model has better denoising performance. Additionally, this model is based on the fact that the noise is spatially independent. Although, the  pixel-shuffle downsampling (PD, stride=2 in DBSN) first introduced in Zhou et al. \cite{zhou2020awgn} are adopted to break the spatial correlation of the noise, the size of the spatial connection that PD (stride=2) can break is finite. For large-scale complex noise in real-world, PD as well as the noise level estimation may not perform well.

AP-BSN \cite{lee2022ap} is a denoising method for sRGB images with real-world noise, proposed by Wooseok Lee et al. in 2022. AP-BSN uses Asymmetric PD (AP) to break the spatial connection of noise, obtains blind spots by masking the central pixel of the convolution kernels, as shown in Fig. \ref{fig:figure 3b}, and adopts random-replacing refinement ($R^3$) to restore image texture details and improve the denoising model's performance.
The stride of PD determines the down-sampling multiple, and different PD strides affect denoising performance during training and testing. The larger the stride, the larger the down-sampling multiple, the more thorough the spatial correlation of the noise is broken, but at the same time the texture information of the image signals is damaged more. AP means adopting PD with different strides during training and testing. AP-BSN uses PD with stride=5 during training and PD with stride=2 during testing for the best performance. 
The AP-BSN loss is the $L_1$-norm distance between the output of the BSN model and the noisy image and can be described as:
\begin{equation}
	\setlength{\abovedisplayskip}{0pt}
	\setlength{\belowdisplayskip}{0pt}
	I_{BSN} = {\rm{PD^{-1}}}(F({\rm{PD}}(I_{N})))
	\label{eq:APBSN_3}
\end{equation}
\begin{equation}
	\setlength{\abovedisplayskip}{0pt}
	\setlength{\belowdisplayskip}{0pt}
	L = \Vert I_{BSN} -I_N \Vert_1
	\label{eq:APBSN_4}
\end{equation}
where $\rm{PD^{-1}}$ means the inverse of PD, $F(\cdot)$ is the BSN model, and $I_{N}$ is the noise image. Taking the noisy image as both input and target. 
AP-BSN is the first denoising algorithm using self-supervised BSN on real-world sRGB images. However, the stride of PD synchronization affects the spatial correlation of noise and the degree of damage to the texture structure of the image signal. Therefore, the stride of PD cannot be too large. AP-BSN's denoising performance may be greatly reduced when the noise in the image is spatially correlated in a large area.

MM-BSN \cite{zhang2023mm} is designed for real-world sRGB images with large areas of spatially correlated noise. MM-BSN is also masked in the network. Unlike AP-BSN \cite{lee2022ap}, which only shields the central pixel of the convolution kernel, MM-BSN uses Multi-mask to shield pixels at multiple positions of the convolution kernel to break the large-area correlation of noise, as shown in Fig. \ref{fig:figure 3c}. 
When the spatial connection area of the noise is large enough, only the central pixel of the convolution kernel is masked, and the surrounding pixels are likely still to be noisy-related. Then, when the feature extraction of the target point is performed through the surrounding pixels, the extracted features will still contain noise. Noise in the real world has different shapes due to different reasons of formation. Therefore, in order to break the spatial correlation of noise, MM-BSN uses masks of various shapes (such as '/', '\verb|\|', '|', '-', '$\square$' etc.) to mask the convolution kernel, the '\verb|\|' shape sample is shown in Fig. \ref{fig:figure 3c}. A noisy image is likely to contain noise of multiple shapes, so it is inappropriate to use only one shape of mask in a model. A novel network structure is proposed to combine the features extracted by Multi-mask. In order for a model to handle noise of various shapes, multiple masks must be used at the same time.
MM-BSN uses the same loss as AP-BSN \cite{lee2022ap}. The noisy image is input to the BSN model in which the pixels at multiple positions of the receptive field are masked for model training.
MM-BSN is the first to propose the feature extraction of images by combining masks of various shapes. The Multi-mask strategy can break the structure of large-area spatial correlation noise and improve the denoising performance of the model. However, the shape of the mask is fixed and cannot be self-adjusted during training according to the shape of the noise.

Li et al. \cite{li2023spatially} proposed the Spatially Adaptive Self-Supervised Learning for Real-World Image Denoising in CVPR2023. It researches how to do the supervision for flat areas and texture areas respectively. For flat areas, it proposed the blind-neighborhood network (BNN), which is deformed from BSN but have different receptive field. At the end of each branch in BNN, the blind spot is created by applying a single pixel shift to the features and a $(2k-1)\times (2k-1)$ blind-neighborhood can be created by increasing the pixel shift size from 1 to $k$.
The loss function for flat areas is:  
\begin{equation}
	\setlength{\abovedisplayskip}{0pt}
	\setlength{\belowdisplayskip}{0pt}
	L_{BNN} = \Vert x_1 -y \Vert_1
	\label{eq:Li_1}
\end{equation}  
$x_1$ is the ouput of BNN, and $y$ is the noisy image.
In addition, to distinguish the texture areas from the flat areas, the paper first calculated the statand deviation map $\sigma$:
\begin{equation}
	\setlength{\abovedisplayskip}{0pt}
	\setlength{\belowdisplayskip}{1pt}
	\begin{aligned}
		\sigma (i, j) &=std(\tilde{x}_1(i-\frac{n-1}{2}:i+\frac{n-1}{2}, \\
		&j-\frac{n-1}{2}:j+\frac{n-1}{2}))
	\end{aligned}
	\label{eq:Li_2}
\end{equation}
where $std(\cdot)$ represents the standard deviation function, which is calculated on 1-channel patches by averaging the values of RGB channels. $n$ is the local window size (set as 7 in this paper).
Then, it proposed the following expression to convert the binarization $\sigma$ to the normalized $\alpha$ with scope of [0, 1]. 
\begin{equation}
	\setlength{\abovedisplayskip}{0pt}
	\setlength{\belowdisplayskip}{0pt}
	\alpha(i,j) = \left\{\begin{array}{ll}
		S(\sigma(i,j)-1), & \sigma(i, j) \leq l\\
		0.5, & l < \sigma(i, j) \leq u\\
		S(\sigma(i,j)-5), & \sigma(i, j) > u
	\end{array} \right.	
	\label{eq:Li_3}
\end{equation}
where $S(\cdot)$ represents the Sigmoid function, $l=1$, and $\mu = 5$. The higher $\alpha (i,j)$ the more local area is textured.
When the texture areas are defined, then a locally aware network (LAN) is presented by stacking $k$ $3\times3$ convolution layers and several $1\times1$ convolution blocks with channel attention mechanism \cite{wen2021openmem}.
The loss function for LAN training phase is as follows:
\begin{equation}
	\setlength{\abovedisplayskip}{0pt}
	\setlength{\belowdisplayskip}{0pt}
	L_{LAN} = (1-\alpha)\cdot \Vert sg(\tilde{x}_1) - \tilde{x}_2 \Vert_1
	\label{eq:Li_4}
\end{equation}
where $\tilde{x}_1$ and $\tilde{x}_2$ denotes the output of the BNN and LAN, respectively. $sg(\cdot)$ represents stop gradient operation \cite{chen2021exploring}.
BNN creates blind spots in the receptive field by shifting the features, which can help to remove noise from flat areas. Although, LAN is specifically designed for texture areas and takes into account the local texture information, this approach may not work well on texture areas where the features are more complex and shifting them may cause loss of important information. Another potential limitation of BNN is that it requires careful tuning of the pixel shift size to create the blind spots, which may be sensitive to the noise characteristics of different images.

\begin{table*}[t]
	\setlength{\abovecaptionskip}{0cm}
	\setlength{\belowcaptionskip}{0cm}
	\caption{Transformer-based self-supervised image denoising methods.}
	\centering
	\begin{tabular}{|m{2.2cm}<{\raggedright}|m{2cm}<{\raggedright}|m{4cm}<{\raggedright}|m{7.5cm}<{\raggedright}|}
		\hline
		Method&	Mask way	&Applications (denoising type)	&Key words (remarks)\\
		\hline
		DT \cite{zhang2023self} &	Mask in inputs	&FM dataset, Gaussian, Poisson and real-world rawRGB images noise denoising.&	CNN and Transformer combined to denoising.\\
		\hline
		LG-BPN \cite{wang2023lg}	&Mask in networks&	Real-world sRGB image noise denoising.&	Densely-Sampled Patch-Masked Convolution (DSPMC) for blind spot get and noise spatial correlated break, Dilated Transformer Block (DTB), combining local information with global information. \\
		\hline
		SwinIA \cite{papkov2023swinia} & $\backslash$ &	FM dataset, Gaussian, Poisson and real-world rawRGB images noise denoising.&	Transformer. \\
		\hline
	\end{tabular}
	\setlength{\belowcaptionskip}{-0.5cm}
	\label{tab:transformer}
\vspace{-2em}
\end{table*}

\subsection{Transformer-based self-supervised image denoising methods}
\noindent Zhang et al. \cite{zhang2023self} proposed a novel self-supervised image denoising method called Denoise Transformer (DT), which takes the masked noisy images as input (as shown in Fig. \ref{fig:figure 2c}) and is constructed using Context-aware Denoise Transformer (CADT) units and a Secondary Noise Extractor (SNE) block. CADT is designed as a dual branch, the local branch composed of several convolutional layers and deformable layers and the global branch composed by transformers. CADTs focus on the fusion and complementarity of local and global features, which boost the denoising performance by residual learning for the noise extraction. What’s more, SNE block is designed in low computational complexity for secondary global noise extraction, simple but effective. DT achieves a competitive performance comparing to the current state-of-the-art methods, especially on denoising images with blurred textures and dark areas. However, the computational complexity of the transformer components is high, which may limit its use in certain industrial applications. Further computational optimization may be needed to make the method more practical for real-world use.

LG-BPN \cite{wang2023lg} also aims to real-world sRGB images denoising based on the architecture combining BSN and Transformer. It fuses local and global information extracted by two branches: a local information extraction branch using multiple cascaded Dilated Convolution Blocks (DCB), and a global information extraction branch using Dilated Transformer Blocks (DTB). To break the noise spatial connection and generate blind spots, LG-BPN uses Densely-Sampled Patch-Masked Convolution (DSPMC) with a specific mask shape for convolution instead of the PD strategy used in AP-BSN \cite{lee2022ap}. Then DCB and DTB are used to extract local and global information.
The mask shape in DSPMC is got from Algorithm \ref{alg1}, where $KH$ and $KW$ denotes the high and wide of convolution kernel, respectively, and mask has a same shape with the convolution kernel. The strategy of replacing the PD with a specific mask convolution kernel can prevent the texture structure of the original image signal from being irreversibly damaged, and allow the subsequent feature extraction to combine more surrounding pixel information, thereby retaining more high-frequency information.
\begin{algorithm}
	\renewcommand{\algorithmicrequire}{\textbf{Input:}}
	\renewcommand{\algorithmicensure}{\textbf{Output:}}
	\caption{Generate the mask}
	\label{alg1}  
	\begin{algorithmic}
		\STATE mask.fill(1)
		\STATE $ dis = KH//2$
		\FOR {$i=0$ to $KH$} 
		\FOR {$j=0$ to $KW$}
		\IF{$abs(i-dis) + abs(j-dis) \leq dis$} 
		\STATE $mask[:, :, i, j]=0$ 
		\ENDIF 
		\ENDFOR
		\ENDFOR 
	\end{algorithmic} 
\end{algorithm}
To maintain the effect of blind spots for DTB, LG-BPN modifies the self-attention calculation and feed-forward layer in the Transformer branch, replacing spatial attention with channel attention and using only $3 \times 3$ dilated convolutions. This allows the Transformer branch to obtain global information while maintaining blind spots, improving denoising performance.
Experimental results on real datasets demonstrate that LG-BPN achieves superior denoising performance, partially resolving the oppositional problem of preserving image texture information and destroying spatial phaseness of real noise through DSPMC and DTB. However, the specific mask shapes in DSPMC may not be optimal for all types of noise patterns, and may not work well on images with complex textures or patterns. Additionally, the modification of the Transformer branch in DTB may increase the computational complexity of the method, which may limit its practical use in certain applications. 

Mikhail et al. \cite{papkov2023swinia} proposed Swin Transformer-based Image Autoencoder (SwinIA), which represents the first convolution-free transformer architecture for blind-spot self-supervised denoising. Unlike other models in its class, this autoencoder does not make any assumptions about the noise distribution and does not rely on masking during training. Additionally, the model can be trained in an autoencoder fashion with a single forward pass and an MSE loss. SwinIA achieves competitive results despite the inherent loss of information from the blind spot and the model's ignorance of the noise nature. However, this method still has the problem of high computational complexity.

\section{Datasets and Evaluation Metrics}
\label{sec:Section3}
\begin{table*}[t]
	\setlength{\abovecaptionskip}{0cm}
	\setlength{\belowcaptionskip}{0cm}
	\caption{Datasets of image denoising.}
	\centering
	\begin{tabular}{m{2.5cm}<{\raggedright}m{2cm}<{\raggedright}m{1.5cm}<{\raggedright}m{2.5cm}<{\raggedright}m{1cm}<{\raggedright}m{1cm}<{\raggedright}m{1cm}<{\raggedright}m{3cm}<{\raggedright}}
		\hline
		Type&Dataset&\multicolumn{2}{c}{Publish}&Color&Greyscale&Set size&Resolution\\
		\hline
		\multirow{13}{*}{Only clean images}&BSD500\cite{arbelaez2010contour}	&2011	&TPAMI&	v	&-	& 500 &	481$\times$321\\
		&DIV2K\cite{agustsson2017ntire}&	2017&	CVPRW&	v	&-	&1000&	1972$\times$1437\\
		&BSD300\cite{martin2001database}&	2001&	ICCV&	v&	-&	500&	481$\times$321\\
		&BSD68\cite{roth2005fields}&	2009&	IJCV&	-&	v&	68&	321$\times$481 \& 481$\times$321\\
		&CBSD68\cite{roth2005fields}	&2005&	-&	v&	-&	68&	321$\times$481 \& 481$\times$321\\
		&Urban100\cite{huang2015single}&	2015&	CVPR&	v&	-&	100&	984$\times$797\\
		&Kodak24\cite{franzen1999kodak}&	1999&	-&	v&	-&	24&	768$\times$512\\
		&Set12\cite{dabov2007image}&	2007&	-&	-&	v&	12&	256$\times$256\\
		&Manga109\cite{aizawa2020building}&	2016&	-&	-&	v&	109&	823$\times$1169\\
		&Set14\cite{zeyde2012single}&	2010&	LNTCS&	v&	-&	14&	492$\times$446\\
		&McMaster\cite{zhang2011color}&	2011&	JEI &	v&	-&	18&	500$\times$500\\
		&Set5\cite{bevilacqua2012low}&	2012&	BMVC&	v&	-&	5&	313$\times$336\\
		&Waterloo Exploration Database\cite{ma2016waterloo}&	2017&	TIP&	v&	-&	4744&	-\\
		\hline
		\multirow{3}{*}{Only noisy images}&	NC12\cite{lebrun2015noise}&	2015&	-&	v&	-	&	- &-\\
		&RNI15\cite{lebrun2015noise}&	2015&	IPOL&	v	&-	&15	&514$\times$465\\
		&RNI6\cite{lebrun2015noise}&	2015&	-	&- &	v&	6&	-\\
		\hline
		\multirow{9}{*}{\makecell[l]{Clean with Real \\noisy images}}&	RENOIR\cite{anaya2018renoir}&	2018&	VCIR&	v&	-&	120	&
		4364$\times$3115\\
		&NAM\cite{nam2016holistic}&	2016&	CVPR&	v&	- &	11&	7360$\times$4912\\
		&CC\cite{nam2016holistic}&	2016&	CVPR&	v&	- &	15&	-\\
		&DND\cite{plotz2017benchmarking}&	2017&	CVPR&	v&	- &	1000&	512$\times$512\\
		&SIDD\cite{abdelhamed2018high}&	2018&	CVPR&	v&	- &	320&	4586$\times$3035\\
		&PolyU\cite{xu2018real}&2018&	arxiv&	v&	- &	100&	2784 $\times$ 1856\\
		&SID\cite{chen2018learning}&	2018&	CVPR&	-&raw &	5094&	5078$\times$3388\\
		&NIND\cite{brummer2019natural}&	2019&	CVPR&	v&	- &	616&	3083$\times$3864\\
		&FMDD\cite{zhang2019poisson}&	2018&	arxiv&	v&	 v &	12,000&	256$\times$256\\
		\hline
	\end{tabular}
	\setlength{\belowcaptionskip}{-0.5cm}
	\label{tab:datasets}
\vspace{-2em}
\end{table*}

\subsection{Datasets}
\noindent In this section, we introduce the widely used public benchmarks for evaluating image denoising performance. These benchmarks include color maps, grayscale maps, and real noise maps, as shown in Table \ref{tab:datasets}, listed according to the published time and image types.

{\bf Kodak24}\cite{franzen1999kodak}: This benchmark consists of 24 high-quality color images with a variety of content and is often used to compare the visual quality of different image processing methods.

{\bf Set5}\cite{bevilacqua2012low}: The Set5 dataset is a collection of 5 images ("baby", "bird", "butterfly", "head", "woman") commonly used for testing the performance of image denoising and image super-resolution models.

{\bf Set12}\cite{dabov2007image}: Set12 is a collection of 12 grayscale images of different scenes that are widely used for evaluating image denoising methods.

{\bf Set14}\cite{zeyde2012single}: The Set14 dataset is a collection of 14 images commonly used for testing visual performance.

{\bf Manga109}\cite{aizawa2020building}: This benchmark is composed of 109 manga volumes drawn by professional manga artists in Japan.

{\bf DIV2K}\cite{agustsson2017ntire}: This is a large dataset of RGB images with a diverse range of contents. The DIV2K dataset contains 800 high-resolution images in the train set, 100 high-resolution images in the validation set, and 100 diverse images in the test set.

{\bf BSD68}\cite{roth2005fields}: This benchmark contains 68 grayscale images and is part of The Berkeley Segmentation Dataset and Benchmark. It is commonly used for measuring the performance of image denoising algorithms.

{\bf BSD300}\cite{martin2001database}: The public benchmark consists of all of the grayscale and color segmentations for 300 images. The images are divided into a training set of 200 images, and a test set of 100 images.

{\bf BSD500} \cite{arbelaez2010contour}: The BSD500 is a dataset of 500 natural images that have been manually annotated with image segmentations. The dataset is commonly used by researchers in the field of computer vision and image processing for evaluating.

{\bf CBSD68}\cite{roth2005fields}: This is the color version of BSD68 and is used for evaluating the performance of color image denoising methods.

{\bf Urban100}\cite{huang2015single}: This is a commonly used test set for evaluating the performance of super-resolution models.

{\bf McMaster} \cite{zhang2011color}: This benchmark contains 18 clean images and is widely used for the test phase.

{\bf RNI6}\cite{lebrun2015noise}: This benchmark contains 6 real noisy grayscale images without ground-truth and is usually used for visualization in the test phase.

{\bf RNI15}\cite{lebrun2015noise}: It contains 15 real noisy color images without ground-truth, and is usually used for visualization in test phase.

{\bf RENOIR}\cite{anaya2018renoir}: It is a dataset of color images corrupted by  natural noise due to low-light conditions, together with spatially and intensity-aligned low noise images of the same scenes.

{\bf CC}\cite{nam2016holistic}: The cc contains 15 real noisy images with different ISOs, i.e., 1,600, 3,200 and 6,400.

{\bf DND}\cite{plotz2017benchmarking}: The DND contained 50 real noisy images and the clean images were captured by low-ISO images.

{\bf Waterloo Exploration Database}\cite{ma2016waterloo}: The Waterloo Exploration Database is a collection of images that have been intentionally degraded to simulate real-world conditions, such as noise, blur, and low resolution. It is used by researchers in the field of image processing to test and compare different algorithms for restoring degraded images to their original quality. The database is available for free and provides a standardized benchmark for evaluating the effectiveness of various image restoration techniques.

{\bf NC12}\cite{lebrun2015noise}: The NC12 contained 12 noisy images and did not have ground-truth clean images. 

{\bf SIDD}\cite{abdelhamed2018high}: The SIDD contained real noisy images from smart phones, and consisted of 320 image pairs of noisy and ground-truth images. 

{\bf NAM}\cite{nam2016holistic}: The Nam includes 11 scenes, which are saved in JPGE format.

{\bf PolyU}\cite{xu2018real}: The PolyU consisted of 100 real noisy images with sizes of $2,784 \times 1, 856$ obtained by five cameras:a Nikon D800, Canon 5D Mark II, Sony A7 II, Canon 80D and Canon 600D.

{\bf SID}\cite{chen2018learning}: The SID(see in the dark) dataset contains 5094 raw short-exposure images, each with a corresponding long-exposure reference image. Images were captured using two cameras: Sony $\alpha 7SII$ and Fujifilm X-T2.

{\bf NIND}\cite{brummer2019natural}: The NIND (Natural Image Noise Dataset) dataset is a set of real photographs with real noise, from identical scenes captured with varying ISO values. Most images are taken with a Fujifilm X-T1 and XF18-55mm, and other photographers are encouraged to contribute images for a more diverse crowdsourced effort.

{\bf FMDD}\cite{zhang2019poisson}: The dataset consists of 12,000 real fluorescence microscopy images obtained with commercial confocal, two-photon, and wide-field microscopes and representative biological samples such as cells, zebrafish, and mouse brain tissues. We use image averaging to effectively obtain ground truth images and 60,000 noisy images with different noise levels.

\subsection{Evaluation Metrics}
\noindent There are several metrics that can be used to evaluate the performance of different denoising algorithms. These include:

{\bf i.} Peak signal-to-noise ratio (PSNR) \cite{hore2010image}: PSNR is measured in decibels (dB) and is commonly used to measure the difference between two images, such as compressed images and original images. It can be used to evaluate the quality of compressed images, restoration of images and ground truth, and the performance of restoration algorithms. A higher PSNR value indicates better image quality.

{\bf ii.} Mean Squared Error (MSE) \cite{wallach1989mean}: MSE is a commonly used metric to evaluate the similarity between two images. It measures the average of the squared differences between the pixel values of the two images, with a lower MSE indicating a higher degree of similarity.

{\bf iii.} Mean Absolute Error (MAE) \cite{willmott2005advantages}: MAE measures the average of the absolute differences between the pixel values of the two images, with a lower MAE indicating a higher degree of similarity. Unlike MSE, MAE gives equal weight to all pixel differences, regardless of their magnitude. The value ranges from 0 to infinity, with smaller values indicating better similarity.

{\bf iv.} Root Mean Square Error (RMSE) \cite{willmott2005advantages}: RMSE measures the square root of the average of the squared differences between the pixel values of the two images, with a lower RMSE indicating a higher degree of similarity. RMSE is similar to MSE, but it is expressed in the same unit as the data being measured, making it easier to interpret. The value ranges from 0 to infinity, with smaller the values indicating better similarity.

{\bf v.} Structural Similarity Index (SSIM) \cite{hore2010image}: SSIM is a full-reference image quality assessment metric that compares the structural information, luminance, and contrast of distorted and reference images. It takes into account the human visual system's sensitivity to these aspects of images and has been shown to provide better correlation with human perception of image quality than other metrics such as MSE. A higher SSIM value indicates better image quality.

{\bf vi.} Figure of Merit (FOM) \cite{balian1977figure}: FOM is a metric used to evaluate the performance of image intensifier tubes used in night vision devices. It takes into account various factors such as resolution, signal-to-noise ratio, and distortion to provide an overall measure of the image quality produced by the image intensifier tube. A higher FOM value indicates better performance.

{\bf vii.} Image Enhancement Factor (IEF) \cite{badamchizadeh2004comparative}: IEF is a measure used to evaluate the effectiveness of image enhancement methods. It is calculated as the ratio of the MSE between the enhanced image and the original image to the MSE between the input image and the original image. A higher IEF value indicates better effectiveness.

{\bf viii.} Feature Similarity Index Mersure(FSIM) \cite{zhang2011fsim}: FSIM is a full-reference image quality assessment metric that compares the similarity of the structural information and luminance of distorted and reference images. It takes into account the human visual system's sensitivity to contrast and spatial frequency and has been shown to correlate well with human perception of image quality.

In this paper, PSNR and SSIM are used for denoising performance comparison, as they are widely used in academic circles to evaluate denoising performance.

The expression of PSNR can be found in following:
\begin{equation}
	\setlength{\abovedisplayskip}{0pt}
	\setlength{\belowdisplayskip}{0pt}
	MSE = \frac{1}{N} \Vert y-x \Vert^2
	\label{eq:MSE}
\end{equation}
\begin{equation}
	\setlength{\abovedisplayskip}{0pt}
	\setlength{\belowdisplayskip}{0pt}
	PSNR= 10 \times \lg (\frac{(2^n -1)^2}{MSE})
	\label{eq:PSNR}
\end{equation}
where $y$ is the denoised results, $x$ is the original image and $N$ is the total number of pixels in the image. $n$ is the number of bits per pixel. 

For two images $x$ and $y$, SSIM calculation can be performed by following:
\begin{equation}
	\setlength{\abovedisplayskip}{0pt}
	\setlength{\belowdisplayskip}{0pt}
	SSIM(x,y) = [L(x,y)]^{\alpha} [C(x,y)]^{\beta} [S(x,y)]^{\gamma}
	\label{eq:SSIM_1}
\end{equation}
here,
\begin{equation}
	\setlength{\abovedisplayskip}{0pt}
	\setlength{\belowdisplayskip}{0pt}
	L(x,y) = \frac{2 \mu_x \mu_y + \epsilon_1}{\mu_x^2+ \mu_y^2 +\epsilon_1}
	\label{eq:SSIM_2}
\end{equation}
\begin{equation}
	\setlength{\abovedisplayskip}{0pt}
	\setlength{\belowdisplayskip}{0pt}
	C(x,y) = \frac{2 \sigma_x \sigma_y + \epsilon_2}{\sigma_x^2 + \sigma_y^2 +\epsilon_2}
	\label{eq:SSIM_3}
\end{equation}
\begin{equation}
	\setlength{\abovedisplayskip}{0pt}
	\setlength{\belowdisplayskip}{0pt}
	S(x,y) = \frac{\sigma_{xy} + \epsilon_3}{\sigma_x \sigma_y +\epsilon_3}
	\label{eq:SSIM_4}
\end{equation}  
where, $\mu_x$ and $\mu_y$ are the mean values, $\sigma_x$ and $\sigma_y$ are the standard deviation, $\sigma_{xy}$ is the covariance of  $x$ and $y$, $\epsilon_1$, $\epsilon_2$ and $\epsilon_3$ are three constants used separately to avoid system errors caused by a denominator of 0.

SSIM models distortion as a combination of three factors: brightness, contrast, and structure. The metric uses the mean as the estimate of brightness, the standard deviation as the estimate of contrast, and the covariance as a measure of structural similarity, satisfying $\alpha >0$, $\beta >0$, $\gamma >0$. The parameters $\alpha$, $\beta$, and $\gamma$ can be used to adjust the relative weight of luminance, contrast, and structural components that contribute to SSIM, respectively.

In general, PSNR and SSIM can provide a quantitative perspective on the denoising quality of images, but it is also important to consider the visual effect comparison, as the human eye perception is sometimes more important.

\section{Experiments}
\label{sec:Section4}
\noindent In this section, we evaluate the performance of recent denoising algorithms on various datasets. We quantitatively and qualitatively compare the denoising performance of the models through the metrics PSNR \cite{hore2010image} and SSIM \cite{hore2010image} and the visual effect of the denoised images. The noise types involved are: Gaussian noise \cite{slepian1962one}, Poisson noise \cite{middleton1951theory}, Impulse noise \cite{henderson1986impulse} and Gamma noise \cite{schultz1964shutdown}. We also evaluate the denoising performance on different types of images, including synthetic noise denoising on grayscale and color images, as well as real-world noise denoising on rawRGB and sRGB images.

\subsection{Synthetic noise denoising on grayscale images}

\begin{figure}[t]
	\setlength{\abovecaptionskip}{0cm}
	\setlength{\belowcaptionskip}{0cm}
	\centering
	\includegraphics[width=1\linewidth]{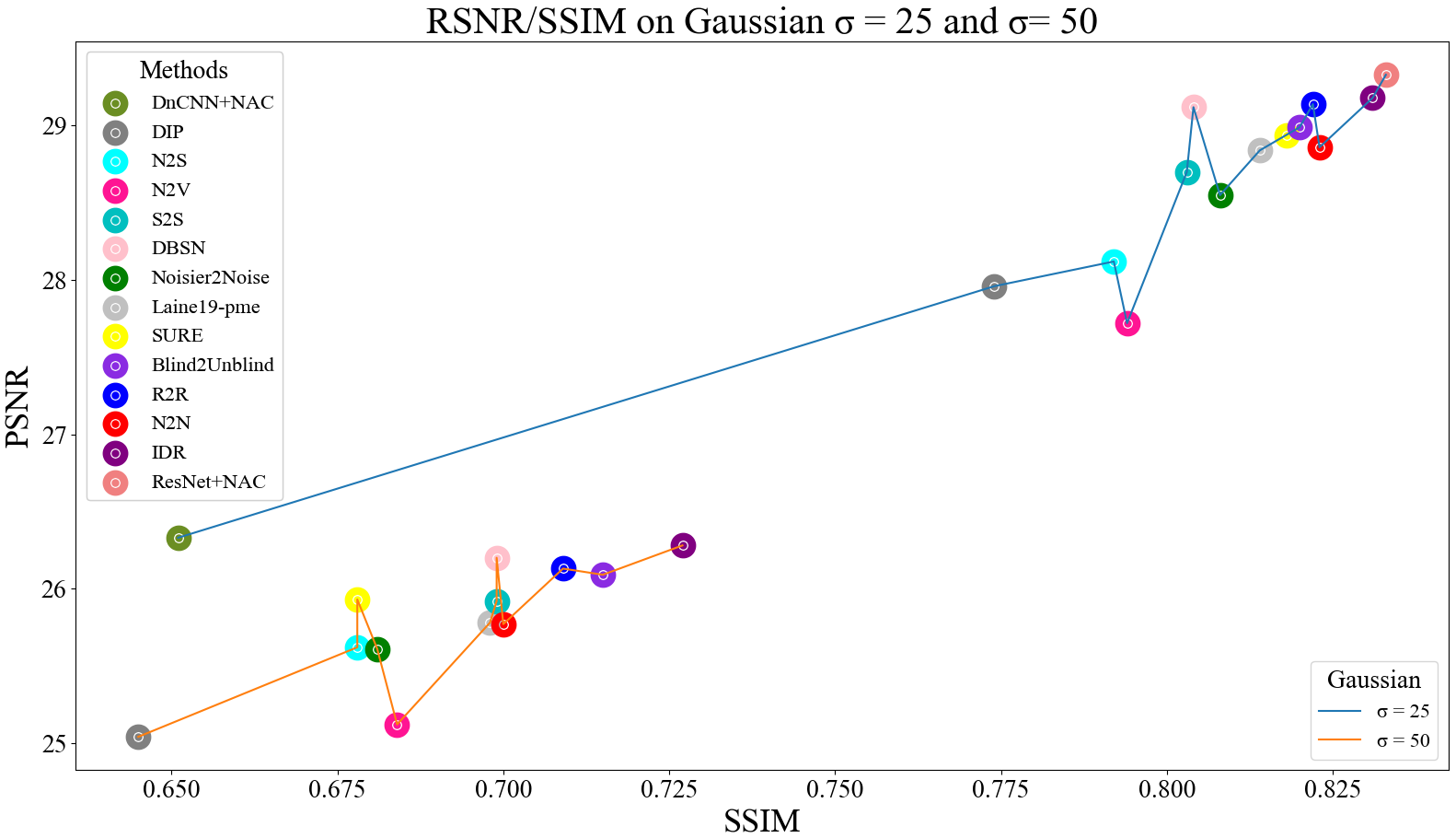}
	\caption{Comparison of different models on denoising BSD68 \cite{roth2005fields} dataset with Gaussian noise of $\sigma$=25 and $\sigma$=50, respectively.}
	\label{fig:figure 4}
\vspace{-1em}
\end{figure}

\begin{figure*}[t]
	\setlength{\abovecaptionskip}{0cm}
	\centering
	\begin{minipage}[b]{0.15\linewidth}
		\centering
		\includegraphics[width=0.7in]{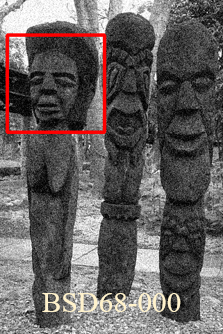}
	\end{minipage}
	\begin{minipage}[b]{0.15\linewidth}
		\centering
		\includegraphics[width=1.04in]{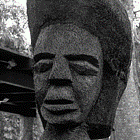}
	\end{minipage}
	\begin{minipage}[b]{0.15\linewidth}
		\centering
		\includegraphics[width=1.04in]{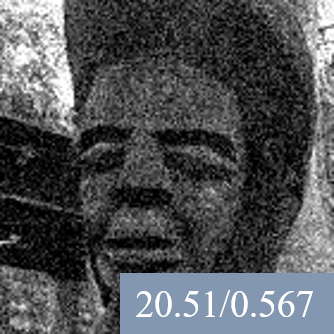}
	\end{minipage}
	\begin{minipage}[b]{0.15\linewidth}
		\centering
		\includegraphics[width=1.04in]{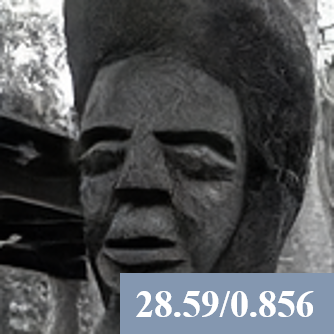}
	\end{minipage}
	\begin{minipage}[b]{0.15\linewidth}
		\centering
		\includegraphics[width=1.04in]{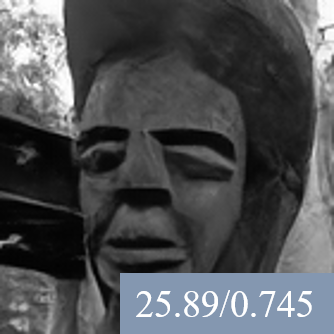}
	\end{minipage}
	\begin{minipage}[b]{0.15\linewidth}
		\centering
		\includegraphics[width=1.04in]{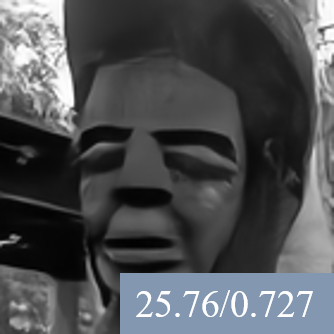}
	\end{minipage}
	
	\vspace{0.06cm}
	\hspace{0.02cm}
	\begin{minipage}[b]{0.15\linewidth}
		\centering
		\subfloat[Original]
		{\includegraphics[width=1.04in]{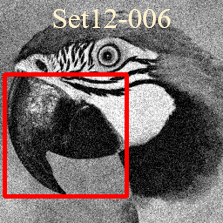}}
	\end{minipage}
	\begin{minipage}[b]{0.15\linewidth}
		\centering
		\subfloat[Clean]
		{\includegraphics[width=1.04in]{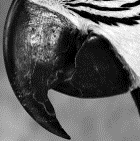}}	
	\end{minipage}
	\begin{minipage}[b]{0.15\linewidth}
		\centering
		\subfloat[Noisy]
		{\includegraphics[width=1.04in]{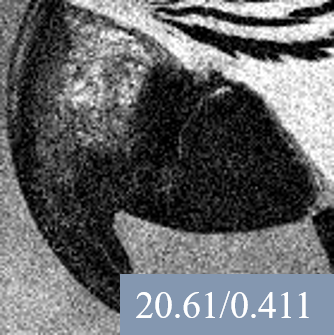}}	
	\end{minipage}
	\begin{minipage}[b]{0.15\linewidth}
		\centering
		\subfloat[Laine et al. \cite{laine2019high}]
		{\includegraphics[width=1.04in]{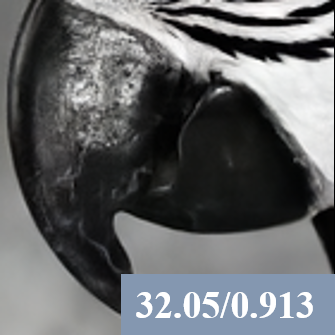}}
	\end{minipage}
	\begin{minipage}[b]{0.15\linewidth}
		\centering
		\subfloat[DBSN \cite{wu2020unpaired}]
		{\includegraphics[width=1.04in]{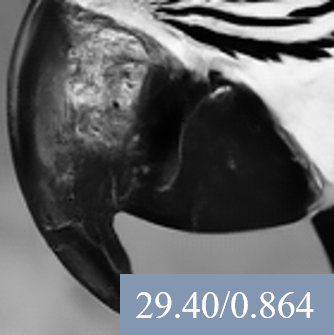}}
	\end{minipage}
	\begin{minipage}[b]{0.15\linewidth}
		\centering
		\subfloat[IDR \cite{zhang2022idr}]
		{\includegraphics[width=1.04in]{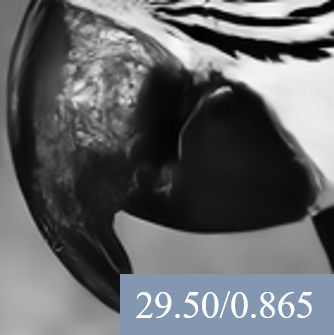}}
	\end{minipage}
	\setlength{\belowcaptionskip}{-0.6cm}
	\caption{Visual Comparison of different methods denoising for grayscale images in BSD68 \cite{roth2005fields} and Set12 \cite{dabov2007image} with Gaussion noise $\sigma=25$.}
	\label{fig:figure gray}
\vspace{-1em}
\end{figure*}

Table \ref{tab:general}-\ref{tab:transformer} show that many models are capable of denoising synthetic noise images. To evaluate the denoising performance of synthetic noise on grayscale images, we take BSD68 \cite{roth2005fields}, and Set12 \cite{dabov2007image} as benchmarks. 

\begin{table}[t]
	\setlength{\abovecaptionskip}{0cm}
	\setlength{\belowcaptionskip}{0cm}
	\caption{The PSNR/SSIM results of Gaussian ($\sigma$=25) denoising on BSD68 \cite{roth2005fields} and Set12 \cite{dabov2007image}. The highest PSNR value is highlighted in \textbf {bold}, the second is \underline{underlined}.}
	\centering
	\begin{tabular*}{\hsize}{c@{\extracolsep{\fill}}c@{\extracolsep{\fill}}c@{\extracolsep{\fill}}}
		\hline
		Method & BSD68 & Set12 \\
		\hline
		N2N \cite{lehtinen2018noise2noise} & 28.86/0.823 & \underline{30.72/0.845}\\
		DIP \cite{ulyanov2018deep} & 27.96/0.774	&25.82/0.772\\
		N2V \cite{krull2019noise2void}	&27.72/0.794&25.01/0.656\\
		GCBD \cite{chen2018image}& 29.15/- & -	\\
		DBSN \cite{wu2020unpaired} & 29.12/0.804 & 30.32/0.835\\
		DnCNN+NAC \cite{xu2020noisy}&	26.33/0.651&	29.21/0.738\\
		ResNet+NAC \cite{xu2020noisy}&	{\bf 29.33/0.833}	&{\bf 31.78/0.880}\\
		S2S \cite{quan2020self2self}&	28.70/0.803& -	\\
		N2S \cite{batson2019noise2self}	&28.12/0.792&	29.16/-\\
		Noise2Score \cite{kim2021noise2score}&	29.12/-&	30.13/-\\
		MC-SURE\cite{ramani2008monte}	&28.94/0.818&	29.13/-\\
		Noisier2Noise \cite{moran2020noisier2noise}&	28.55/0.808	& -\\
		Laine19-pme \cite{laine2019high}	&28.84/0.814&	\\
		R2R\cite{pang2021recorrupted}	&29.14/0.822&	30.06/0.851\\
		B2UB\cite{wang2022blind2unblind}&	28.99/0.820	&30.09/0.854\\
		IDR \cite{zhang2022idr}&	\underline{29.20/0.835}	&30.40/0.863\\
		\hline
	\end{tabular*}
	\setlength{\belowcaptionskip}{-0.5cm}
	\label{tab:gray_g_25}
\vspace{-2em}
\end{table}

\begin{table}[t]
	\setlength{\abovecaptionskip}{0cm}
	\setlength{\belowcaptionskip}{0cm}
	\caption{The PSNR/SSIM results of Gaussian ($\sigma$=50) denoising on BSD68 \cite{roth2005fields} and Set12 \cite{dabov2007image}. The highest PSNR value is highlighted in \textbf {bold}, the second is \underline{underlined}.}
	\centering
	\begin{tabular*}{\hsize}{c@{\extracolsep{\fill}}c@{\extracolsep{\fill}}c@{\extracolsep{\fill}}}
		\hline
		Method & BSD68 & Set12 \\
		\hline
		N2N \cite{lehtinen2018noise2noise}	 &25.77/0.700 &	\underline{27.20/-}\\
		DIP \cite{ulyanov2018deep} &	25.04/0.645 & -	\\
		N2V \cite{krull2019noise2void}	 &25.12/0.684 &	24.68/-\\
		DBSN \cite{wu2020unpaired} &	26.20/0.699	 &27.19/0.758\\
		S2S \cite{quan2020self2self} &	25.92/0.699	 & -\\
		N2S \cite{batson2019noise2self}	 &25.62/0.678 &	26.19/-\\
		Noise2Score \cite{kim2021noise2score} &	26.21/- &	27.16/-\\
		MC-SURE \cite{ramani2008monte} &	25.93/0.678 &	26.23/-\\
		Noisier2Noise \cite{moran2020noisier2noise} &	25.61/0.681	 & -\\
		Laine19-pme \cite{laine2019high}	 &25.78/0.698 & -	\\
		R2R\cite{pang2021recorrupted}	 &\underline{26.13/0.709} &	26.86/0.771\\
		B2UB\cite{wang2022blind2unblind} &	26.09/0.715	 &26.91/0.776\\
		IDR \cite{zhang2022idr} &	{\bf 26.25/0.726}	 & {\bf 27.29/0.789}\\
		\hline
	\end{tabular*}
	\setlength{\belowcaptionskip}{-0.5cm}
	\label{tab:gray_g_50}
\vspace{-2em}
\end{table}

Table \ref{tab:gray_g_25} and Table \ref{tab:gray_g_50} show PSNR/SSIM results of different models on BSD68\cite{roth2005fields} and Set12\cite{dabov2007image} datasets for removing Gaussian noise with levels of $\sigma$=25 and $\sigma$=50. The tables reveal that the model with the best denoising effect and the worst model are not always the same on different datasets. When $\sigma$ = 25, N2N \cite{lehtinen2018noise2noise}, ResNet+NAC \cite{xu2020noisy}, and IDR \cite{zhang2022idr} are among the best performing models, while when $\sigma$ = 50, IDR performs the best because its Iterative Data Refinement strategy is effective in removing high-level noise. However, these models require real noisy-noisy image pairs or noise models in addition to single noise images.

Since most models are evaluated on BSD68\cite{roth2005fields}, we draw curves in Fig. \ref{fig:figure 4} to visually compare the denoising performance of different models at different noise levels. As shown in Fig. \ref{fig:figure 4}, all models have much higher denoising performance with a noise level of 25 compared to the performance with a noise level of 50. This indicates that the larger the noise level, the more severe the damage to the details of the image, and the more challenging it becomes to strike a balance between restoring details and removing noise. 

Fig. \ref{fig:figure gray} presents the comparison of visual performance by self-supervised Laine et al. \cite{laine2019high}, DBSN \cite{wu2020unpaired} and IDR \cite{zhang2022idr} denoising on the grayscale images BSD68 \cite{roth2005fields} and Set12 \cite{dabov2007image} datasets. Laine et al. achieves the best denoising performance on grayscale images, preserving more details during denoising. However, the image denoised by IDR appears too smooth, which is an inevitable result of the iterative method used for model optimization.

\subsection{Synthetic noise denoising on color images}

To evaluate the denoising performance of synthetic noise on grayscale images, we show the PSNR/SSIM results of different methods on several datasets such as Kodak \cite{franzen1999kodak}, CBSD68 \cite{roth2005fields}, BSD300 \cite{martin2001database}, Set9 \cite{ulyanov2018deep}, and Set14 \cite{zeyde2012single}. The noise types covered in this section include Gaussian, Poisson, Impulse, Bernoulli and Gamma noise. While a large number of models use Gaussian and Poisson noise for experiments, only a few studies evaluate the effectiveness of removing other types of noise.

Table \ref{tab:color_impulse} displays the Impulse noise removal results on the Kodak\cite{franzen1999kodak}, BSD300\cite{martin2001database}, and Set14\cite{zeyde2012single} datasets. The results show that Laine19-pme \cite{laine2019high} achieves a comparable effect to N2N \cite{lehtinen2018noise2noise} in the removal of Impulse noise. Additionally, the noise removal ability of models for fixed noise levels is higher than that of noise levels that vary within a certain range.

\begin{figure*}[t]
	\setlength{\abovecaptionskip}{-0.2cm}
	\begin{minipage}{0.332\linewidth}
		\centering
		\subfloat[]
		{\includegraphics[height=1.35in]{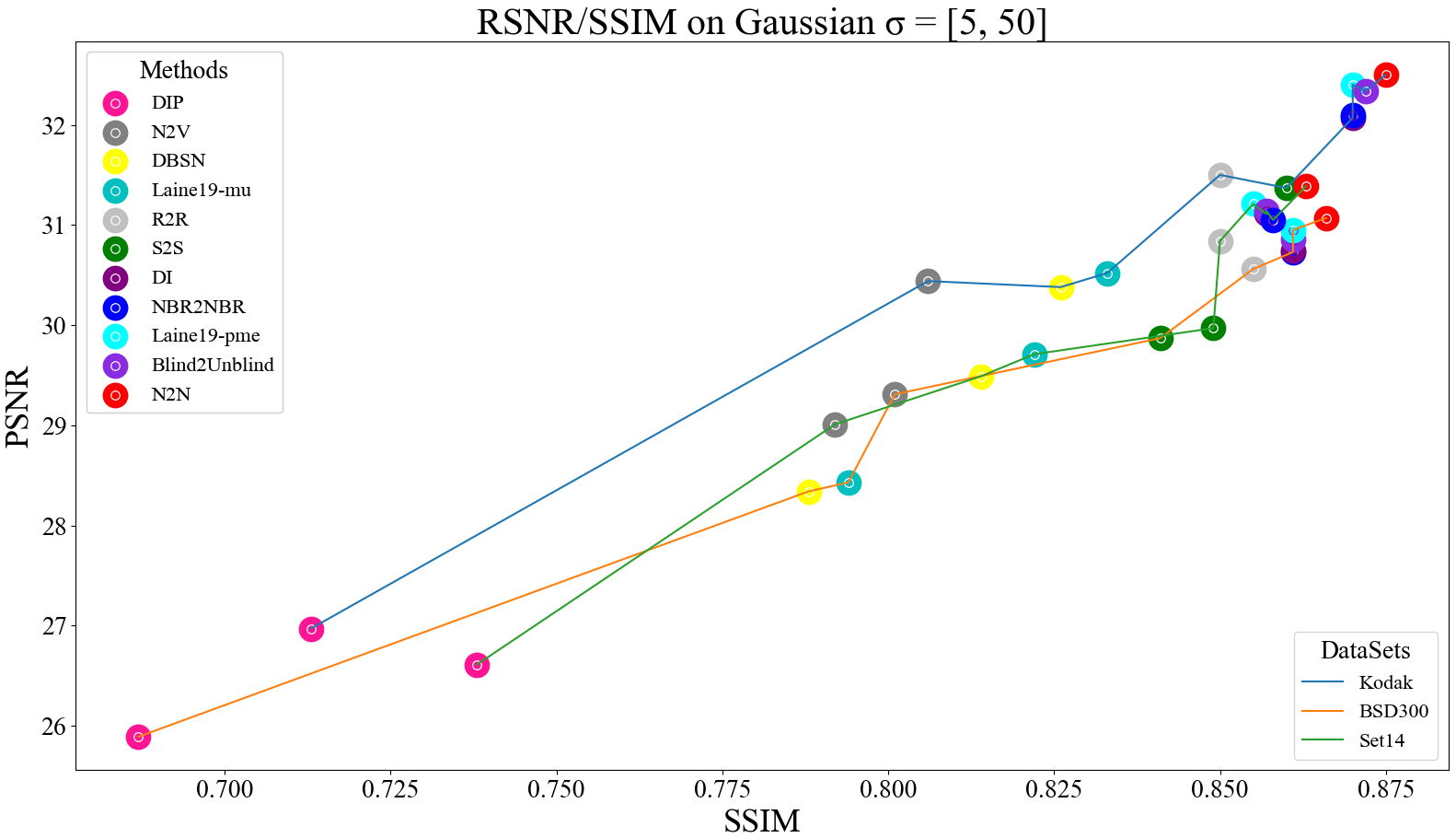}}
	\end{minipage}
	\begin{minipage}{0.332\linewidth}
		\centering
		\subfloat[]
		{\includegraphics[height=1.35in]{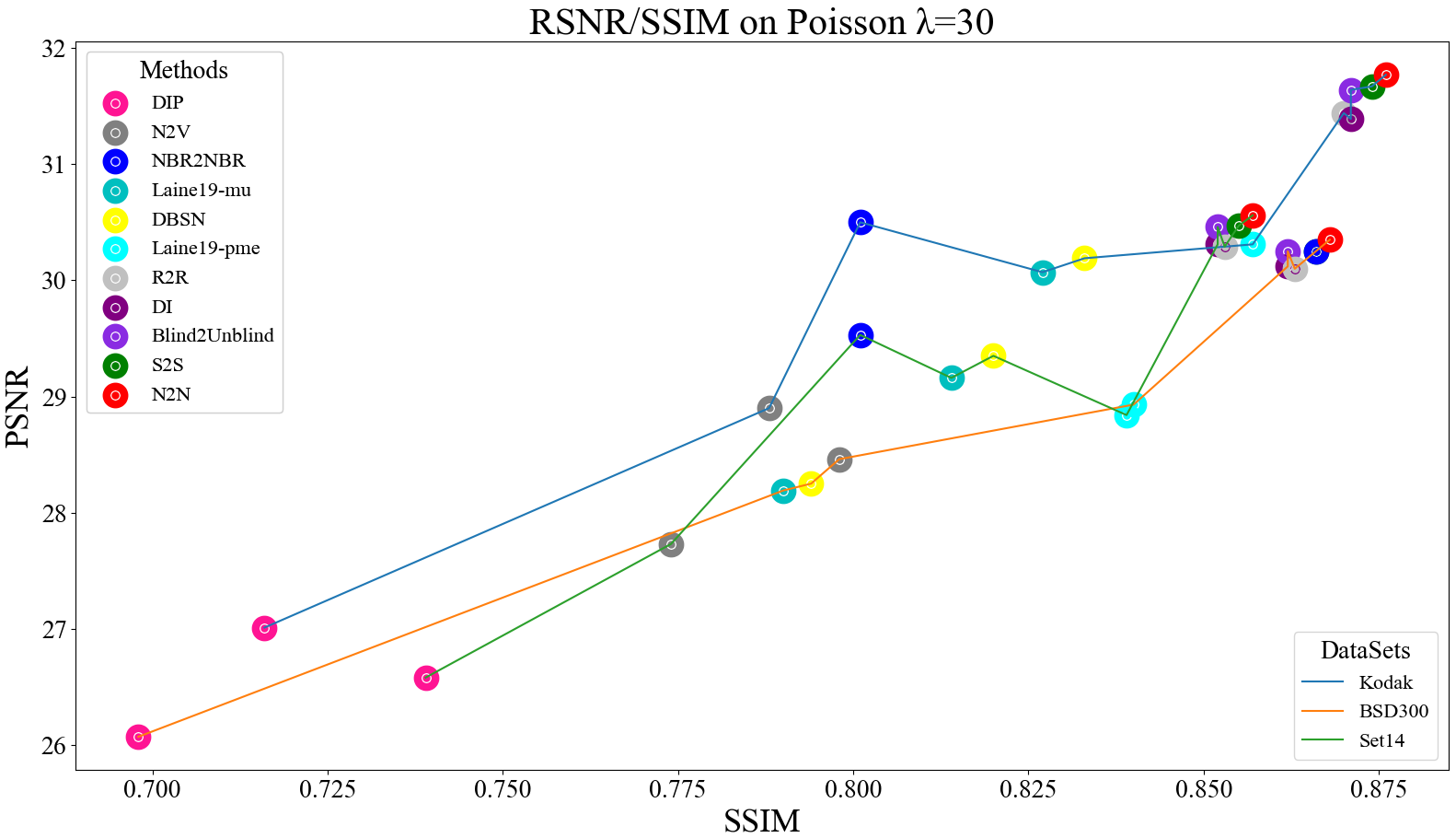}}
	\end{minipage}
	\begin{minipage}{0.332\linewidth}
		\centering
		\subfloat[]
		{\includegraphics[height=1.35in]{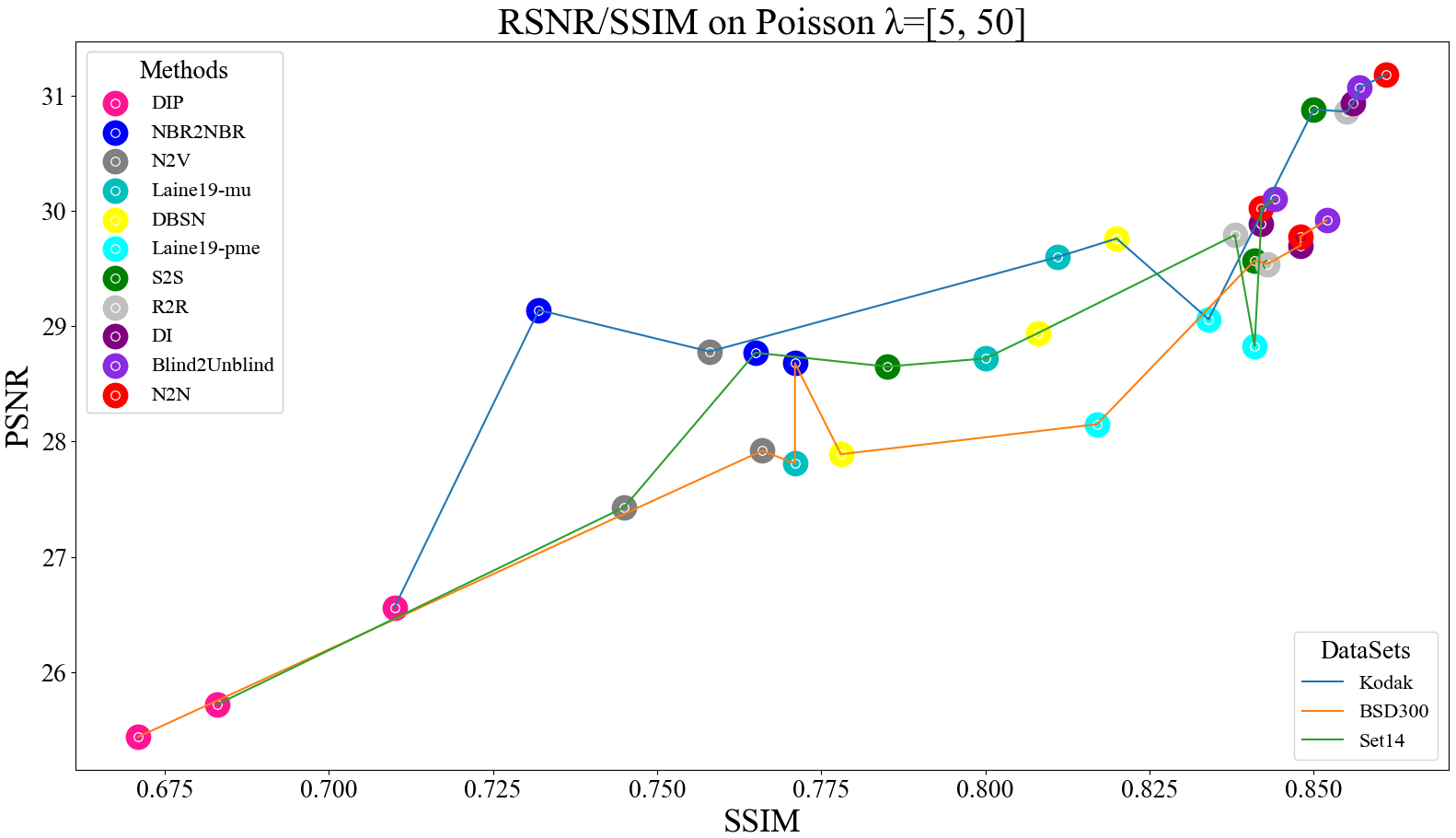}}
	\end{minipage}
	\setlength{\belowcaptionskip}{-0.3cm}
	\caption{Comparison of different models on Gaussian ($\sigma$=[5, 50]) noise, Possion noise ($\lambda$=30, and $\lambda$=[5, 50]) denoising on Kodak \cite{franzen1999kodak}, BSD300 \cite{martin2001database} and Set14 \cite{zeyde2012single}.}
	\label{fig:figure 6}
\vspace{-2em}
\end{figure*}

\begin{table}[t]
	\setlength{\abovecaptionskip}{0cm}
	\setlength{\belowcaptionskip}{0cm}
	\caption{The PSNR results of Impulse noise denoising on Kodak \cite{franzen1999kodak}, BSD300 \cite{martin2001database} and Set14 \cite{zeyde2012single}.}
	\centering
	\begin{tabular*}{\hsize}{c@{\extracolsep{\fill}}c@{\extracolsep{\fill}}c@{\extracolsep{\fill}}c@{\extracolsep{\fill}}c@{\extracolsep{\fill}}}
		\hline
		Noisy type & Method & Kodak&	BSD300&	Set14\\
		\hline
		\multirow{3}{*}{Impulse ($\alpha$=0.5)}&	N2N \cite{lehtinen2018noise2noise}	&32.88	&30.85&	30.94\\
		&Laine19-mu \cite{krull2019noise2void}	&30.82&	28.52&	29.05\\
		&Laine19-pme \cite{laine2019high}&	32.93&	30.71&	31.09\\
		\hline
		\multirow{3}{*}{Impulse ($\alpha \in$[0,1])}&	N2N \cite{lehtinen2018noise2noise}&	31.53&	30.11&	29.51\\
		&Laine19-mu \cite{laine2019high}&	27.16&	25.55&	25.56\\
		&Laine19-pme \cite{laine2019high}&	31.40&	29.98&	29.51\\
		\hline
	\end{tabular*}
	\setlength{\belowcaptionskip}{-0.5cm}
	\label{tab:color_impulse}
\vspace{-2em}
\end{table}

Table \ref{tab:color_gamma} displays the denoising performance of N2N \cite{lehtinen2018noise2noise}, N2V \cite{krull2019noise2void}, N2S \cite{batson2019noise2self}, Noise2Score \cite{kim2021noise2score}, NBR2NBR \cite{huang2021neighbor2neighbor} and Kim et al. \cite{kim2022noise} on Gamma noise. The table shows that N2N performs best to remove Gamma noise on all datasets. Additionally, the denoising performance of different methods at different noise levels is consistent.

\begin{table}[t]
	\setlength{\abovecaptionskip}{0cm}
	\setlength{\belowcaptionskip}{0cm}
	\caption{The PSNR results of Gamma noise denoising on Kodak \cite{franzen1999kodak}, CBSD68 \cite{roth2005fields}.}
	\centering
	\begin{tabular*}{\hsize}{c@{\extracolsep{\fill}}c@{\extracolsep{\fill}}c@{\extracolsep{\fill}}c@{\extracolsep{\fill}}}
		\hline
		Noisy type & Method & Kodak&	CBSD68\\
		\hline
		\multirow{6}{*}{Gamma (K=100)}&	N2N \cite{lehtinen2018noise2noise}&	36.26&	35.45\\
		&N2V \cite{krull2019noise2void}&	31.96&	31.83\\
		&N2S \cite{batson2019noise2self}&	32.83&	31.71\\
		&Noise2Score \cite{kim2021noise2score}&	34.23&	33.82\\
		&NBR2NBR \cite{huang2021neighbor2neighbor}&	35.10&	34.21\\
		&Kim et al. \cite{kim2022noise}&	35.42&	34.52\\
		\hline
		\multirow{6}{*}{Gamma (K=50)}&	N2N \cite{lehtinen2018noise2noise}&	34.49&	33.52\\
		&N2V \cite{krull2019noise2void}&	31.38&	30.51\\
		&N2S \cite{batson2019noise2self}	&31.71&	30.63\\
		&Noise2Score \cite{kim2021noise2score}&	31.34&	31.05\\
		&NBR2NBR \cite{huang2021neighbor2neighbor}&	32.38&	32.11\\
		&Kim et al. \cite{kim2022noise}&	32.81&	32.43\\
		\hline
	\end{tabular*}
	\setlength{\belowcaptionskip}{-0.5cm}
	\label{tab:color_gamma}
\vspace{-2em}
\end{table}

Table \ref{tab:color_g_25} and Table \ref{tab:color_g_50} evaluate the denoising ability of more than a dozen self-supervised denoising methods considering Gaussian noise with a noise level of 25 and 50 on Kodak\cite{franzen1999kodak}, CBSD68\cite{roth2005fields}, BSD300\cite{roth2005fields}, Set9\cite{ulyanov2018deep}, and Set14\cite{zeyde2012single} datasets using PSNR/SSIM. The tables show that on Kodak and Set14, N2N is the best among all models, and it ahieves the second-best denoising on CBSD68, BSD300, and Set9. Overall, when $\sigma$ = 25, N2N\cite{lehtinen2018noise2noise} achieves the best denoising effect, while there exist methods that surpass N2N\cite{lehtinen2018noise2noise} in some datasets. This finding suggests that the denoising method using single noise images can achieve denoising performance comparable to or even better than that of N2N, which requires real noisy-noisy image pairs. When $\sigma$ = 50, IDR \cite{zhang2022idr} achieved the best performance for its effectiveness in removing high-level noise, which is consistent with the performance on grayscale images denoising.

\begin{table*}[t]
	\setlength{\abovecaptionskip}{0pt}
	\setlength{\belowcaptionskip}{0pt}
	\caption{The PSNR/SSIM results of Gaussian ($\sigma$ = 25) noise denoising on Kodak \cite{franzen1999kodak}, CBSD68 \cite{roth2005fields}, BSD300 \cite{martin2001database}, Set9 \cite{ulyanov2018deep}, and Set14 \cite{zeyde2012single}. The highest PSNR value is highlighted in \textbf {bold}, the second is \underline{underlined}.}
	\centering
	\begin{tabular*}{\hsize}{c@{\extracolsep{\fill}}l@{\extracolsep{\fill}}l@{\extracolsep{\fill}}l@{\extracolsep{\fill}}l@{\extracolsep{\fill}}l@{\extracolsep{\fill}}}
		\hline
		Method	&Kodak&	CBSD68&	BSD300&	Set9&	Set14\\
		\hline
		N2N \cite{lehtinen2018noise2noise}	&{\bf 32.41/0.884}&	\underline{31.10/-}&	\underline{31.39/0.889}&	\underline{31.33/0.957}&	{\bf 31.37/0.868}\\
		MC-SURE \cite{ramani2008monte}&	30.75/-	&30.23/-&	-&	-&	-\\
		DIP \cite{ulyanov2018deep}&	27.20/0.720&	-&	26.38/0.708&	30.77/0.942&	27.16/0.758\\
		N2V \cite{krull2019noise2void}	&31.63/0.869&	29.22/-&	30.72/0.874&	30.66/0.947&	28.84/0.802\\
		N2S \cite{batson2019noise2self}	&30.81/-&	30.05/-&	-&	30.05/0.944&	-\\
		Laine19-mu \cite{laine2019high}&	30.62/0.840&	28.61/-&	28.62/0.803&	-&	29.93/0.830\\
		Laine19-pme \cite{laine2019high}&	\underline{32.40/0.883}&	30.88/-&	30.99/0.877&	-&	\underline{31.36/0.866}\\
		DBSN \cite{wu2020unpaired}&	32.07/0.875&	-&	31.12/0.881&	-&	30.63/0.846\\
		Noisier2Noise \cite{moran2020noisier2noise}&	31.96/0.869&	-&	29.57/0.815&	-&	29.64/0.832\\
		S2S \cite{quan2020self2self}&	31.28/0.864&	-&	29.86/0.849&	{\bf 31.74/0.956}&	30.08/0.839\\
		Noise2Score \cite{kim2021noise2score}&	31.89/-&	30.85/-&	-&	-&	-\\
		R2R \cite{pang2021recorrupted}&	32.25/0.880&	-&	30.91/0.872&	-&	31.32/0.865\\
		NBR2NBR \cite{huang2021neighbor2neighbor}&	32.08/0.879&	30.56/-&	30.79/0.873&	-&	31.09/0.864\\
		IDR \cite{zhang2022idr}&	32.36/0.884&	{\bf 31.29/0.889}&	{\bf 31.48/0.890}&	-&	30.85/0.866\\
		Kim et al. \cite{kim2022noise}&	31.78/-&	30.89/-&	-&	-&	-\\
		B2UB\cite{wang2022blind2unblind}	&32.27/0.880&	-	&30.87/0.872	&-	&31.27/0.864\\
		DT \cite{zhang2023self}	&31.96/0.879&	-&	30.80/0.872	&-	&31.15/0.864\\
		\hline
	\end{tabular*}
	\setlength{\belowcaptionskip}{-0.2cm}
	\label{tab:color_g_25}
\vspace{-1em}
\end{table*}

\begin{table}[t]
	\setlength{\abovecaptionskip}{0pt}
	\setlength{\belowcaptionskip}{0pt}
	\caption{The PSNR/SSIM results of Gaussian ($\sigma$ = 50) noise denoising on Kodak \cite{franzen1999kodak}, CBSD68 \cite{roth2005fields}, BSD300 \cite{martin2001database} and Set9 \cite{ulyanov2018deep}. The highest PSNR value is highlighted in \textbf {bold}, the second is \underline{underlined}.}
	\centering
	\begin{tabular*}{\hsize}{c@{\extracolsep{\fill}}l@{\extracolsep{\fill}}l@{\extracolsep{\fill}}l@{\extracolsep{\fill}}l@{\extracolsep{\fill}}}
		\hline
		Method	&Kodak&	CBSD68&	BSD300&	Set9\\
		\hline
		N2N \cite{lehtinen2018noise2noise}	&\underline{29.23/0.803}&	\underline{27.94/-}&	\underline{28.17/0.799}&	\underline{28.94/0.929}\\
		MC-SURE\cite{ramani2008monte}	&26.93/-&	26.24/-&	-&	-\\
		DIP \cite{ulyanov2018deep}&	-&	-&	-	&28.23/0.910\\
		N2V \cite{krull2019noise2void}	&28.57/0.776&	25.13/-&	27.60/0.775&	27.81/0.912\\
		N2S \cite{batson2019noise2self}&	28.21/-	&27.14/-	&-	&27.51/0.905\\
		Laine19-mu \cite{laine2019high}&	27.78/-&	26.42/-&	-&	-\\
		Laine19-pme \cite{laine2019high}	&28.63/-&	27.65/-	&-	&-\\
		DBSN \cite{wu2020unpaired}	&28.81/0.783	&-	&27.87/0.782	&-\\
		Noisier2Noise \cite{moran2020noisier2noise}&	28.73/0.770&	-&	26.18/0.684	&-\\
		S2S \cite{quan2020self2self}	&-	&-	&-	&{\bf 29.25/0.928}\\
		Noise2Score \cite{kim2021noise2score}&	28.83/-	&27.75/-&	-&	-\\
		NBR2NBR \cite{huang2021neighbor2neighbor}	&28.28/-	&27.32/-	&-&	-\\
		IDR \cite{zhang2022idr}	&{\bf 29.27/0.803}&	{\bf 28.09/0.800}& {\bf 28.25/0.802} &-\\
		Kim et al. \cite{kim2022noise}	&28.64/-&	27.56/-	&-&	-\\
		\hline
	\end{tabular*}
	\setlength{\belowcaptionskip}{-0.5cm}
	\label{tab:color_g_50}
\vspace{-1em}
\end{table}

\begin{figure*}[t]
	\setlength{\abovecaptionskip}{0cm}
	\setlength{\belowcaptionskip}{0cm}
	\centering
	\begin{minipage}[b]{0.24\linewidth}
		\setlength{\abovecaptionskip}{1pt}
		\centering
		\captionsetup[subfloat]{labelsep=none,format=plain,labelformat=empty, font={scriptsize}}
		\subfloat[BSD300-087]{\includegraphics[width=1.7in]{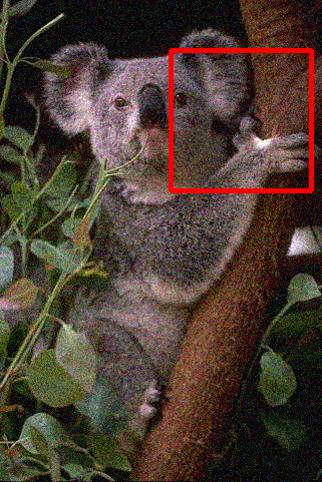}}	
	\end{minipage}
	\begin{minipage}[b]{0.16\linewidth}
		\setlength{\abovecaptionskip}{1pt}
		\centering
		\addtocounter{subfigure}{-1}
		\captionsetup{font={scriptsize}}
		\subfloat[Clean]{\includegraphics[height=1.119in]{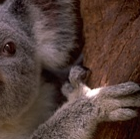}}
		\quad
		\captionsetup{font={scriptsize}}
		\subfloat[Noisy]{\includegraphics[width=1.12in]{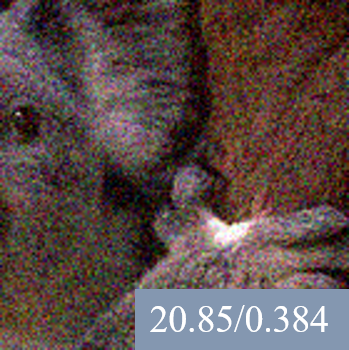}}
	\end{minipage}
	\begin{minipage}[b]{0.16\linewidth}
		\setlength{\abovecaptionskip}{1pt}
		\centering
		\captionsetup{font={scriptsize}}
		\subfloat[N2N \cite{lehtinen2018noise2noise}]
		{\includegraphics[height=1.119in]{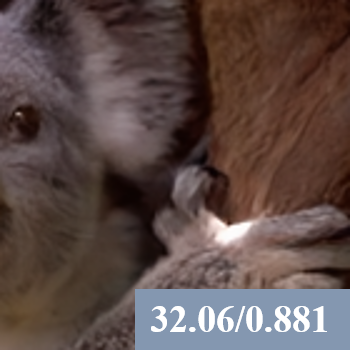}}
		\quad
		\captionsetup{font={scriptsize}}
		\subfloat[Laine et al. \cite{laine2019high}]
		{\includegraphics[width=1.12in]{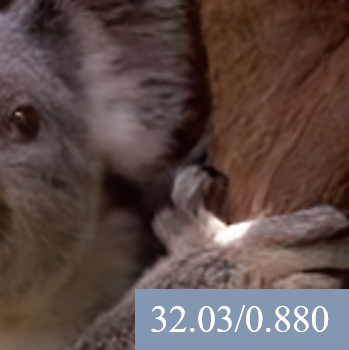}}
	\end{minipage}
	\begin{minipage}[b]{0.16\linewidth}
		\setlength{\abovecaptionskip}{1pt}
		\centering
		\captionsetup{font={scriptsize}}
		\subfloat[NBR2NBR \cite{huang2021neighbor2neighbor}]
		{\includegraphics[height=1.119in]{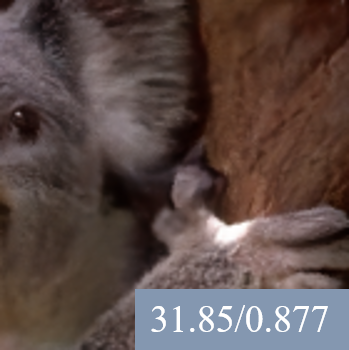}}
		\quad
		\captionsetup{font={scriptsize}}
		\subfloat[IDR \cite{zhang2022idr}]
		{\includegraphics[width=1.12in]{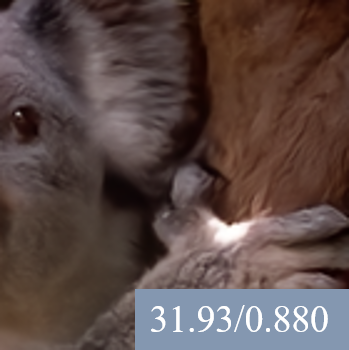}}
	\end{minipage}
	\begin{minipage}[b]{0.16\linewidth}
		\setlength{\abovecaptionskip}{1pt}
		\centering
		\captionsetup{font={scriptsize}}
		\subfloat[B2UB \cite{wang2022blind2unblind}]
		{\includegraphics[height=1.119in]{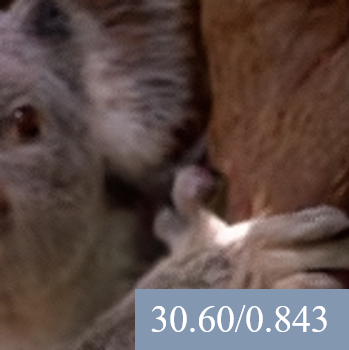}}
		\captionsetup{font={scriptsize}}
		\subfloat[DT\cite{zhang2023self}]
		{\includegraphics[width=1.12in]{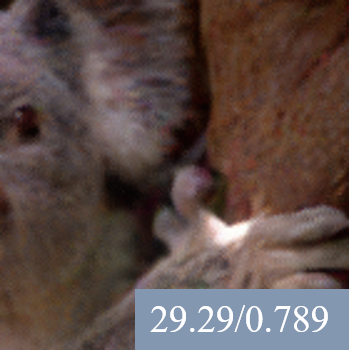}}
	\end{minipage}

	\setlength{\belowcaptionskip}{-0.3cm}
	\caption{Visual comparison of different models denoising for BSD300-087 \cite{martin2001database} image with Gaussian noise $\sigma$=25.}
	\label{fig:figure color_bsd300}
\end{figure*}

\begin{figure*}[t]
	\setlength{\abovecaptionskip}{0cm}
	\setlength{\belowcaptionskip}{0cm}
	\centering
	\begin{minipage}[b]{0.44\linewidth}
		\setlength{\abovecaptionskip}{1pt}
		\centering
		\captionsetup[subfloat]{labelsep=none,format=plain,labelformat=empty, font={scriptsize}}
		\subfloat[Kodak24-006]{\includegraphics[height=2.105in]{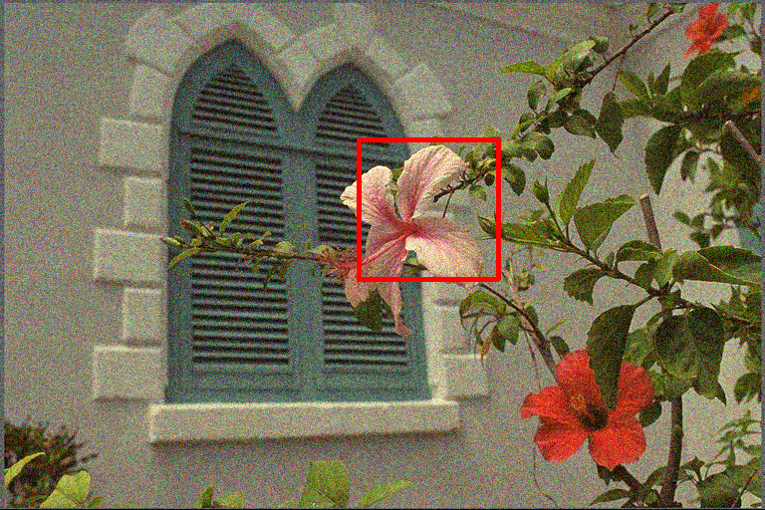}}	
	\end{minipage}
	\begin{minipage}[b]{0.13\linewidth}
		\setlength{\abovecaptionskip}{1pt}
		\centering
		\addtocounter{subfigure}{-1}
		\captionsetup{font={scriptsize}}
		\subfloat[Clean]{\includegraphics[height=0.9in]{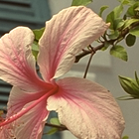}}
		\quad
		\captionsetup{font={scriptsize}}
		\subfloat[Noisy]{\includegraphics[height=0.9in]{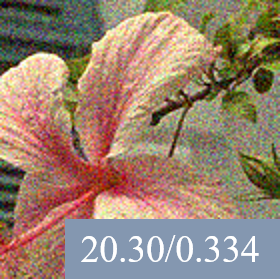}}
	\end{minipage}
	\begin{minipage}[b]{0.13\linewidth}
		\setlength{\abovecaptionskip}{1pt}
		\centering
		\captionsetup{font={scriptsize}}
		\subfloat[N2N \cite{lehtinen2018noise2noise}]
		{\includegraphics[height=0.9in]{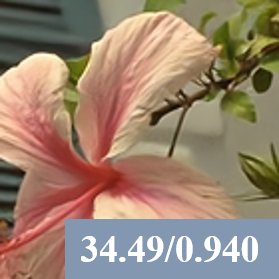}}
		\quad
		\captionsetup{font={scriptsize}}
		\subfloat[Laine et al. \cite{laine2019high}]
		{\includegraphics[height=0.9in]{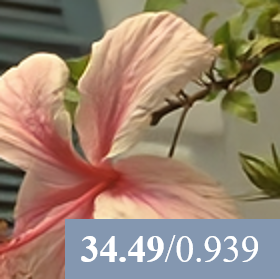}}
	\end{minipage}
	\begin{minipage}[b]{0.13\linewidth}
		\setlength{\abovecaptionskip}{1pt}
		\centering
		\captionsetup{font={scriptsize}}
		\subfloat[NBR2NBR \cite{huang2021neighbor2neighbor}]
		{\includegraphics[height=0.9in]{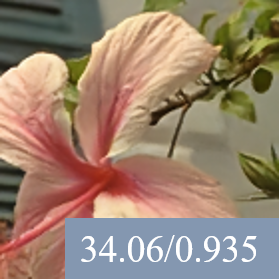}}
		\quad
		\captionsetup{font={scriptsize}}
		\subfloat[IDR \cite{zhang2022idr}]
		{\includegraphics[height=0.9in]{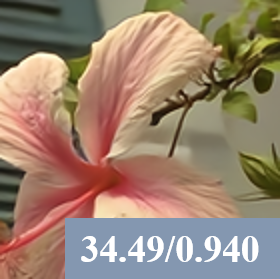}}
	\end{minipage}
	\begin{minipage}[b]{0.13\linewidth}
		\setlength{\abovecaptionskip}{1pt}
		\centering
		\captionsetup{font={scriptsize}}
		\subfloat[B2UB \cite{wang2022blind2unblind}]
		{\includegraphics[height=0.9in]{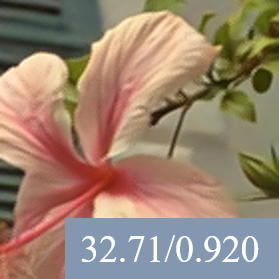}}
		\captionsetup{font={scriptsize}}
		\subfloat[DT\cite{zhang2023self}]
		{\includegraphics[height=0.9in]{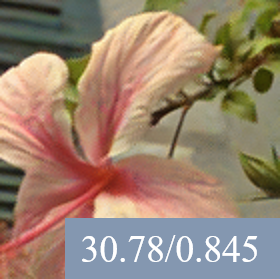}}
	\end{minipage}
	
	\setlength{\belowcaptionskip}{-0.6cm}
	\caption{Visual comparison of different models denoising the Kodak24-006 \cite{franzen1999kodak} image with Gaussian noise $\sigma$=25.}
	\label{fig:figure color_kodak}
\vspace{-1em}
\end{figure*}

Fig. \ref{fig:figure 6} shows the denoising performance of 11 self-supervised image denoising models on the Kodak\cite{franzen1999kodak}, BSD300\cite{roth2005fields} and Set14\cite{zeyde2012single} datasets for Gaussian noise and Poisson noise. When the Gaussian noise level fluctuates from 5 to 50, N2N \cite{lehtinen2018noise2noise} and DIP \cite{ulyanov2018deep} are the best and worst models for denoising, respectively, and the denoising performance of Laine19-pme \cite{laine2019high}, B2UB \cite{wang2022blind2unblind}, DT \cite{zhang2023self} and NBR2NBR \cite{huang2021neighbor2neighbor} are close to that of N2N. When the Poisson noise level is fixed at 30, N2N and DIP are also the methods with the best and worst denoising effects, respectively, and the denoising performance of S2S \cite{quan2020self2self}, B2UB, DT, and R2R \cite{pang2021recorrupted} approach that of N2N. When the Poisson noise level fluctuates from 5 to 50, B2UB achieves the best denoising performance on BSD300 and Set14, except that N2N remains the best denoising effect on the Kodak dataset. In this case, the denoising performance of DT and R2R are closely followed by N2N and B2UB. It is worth noting that most of the methods that are close to or even exceed the denoising performance of N2N are BSN-based models. This observation suggests that BSN-based models have great potential for development.

We also show the visual performance of self-supervised models N2N \cite{lehtinen2018noise2noise}, Laine et al. \cite{laine2019high}, NBR2NBR \cite{huang2021neighbor2neighbor}, IDR \cite{zhang2022idr}, B2UB \cite{wang2022blind2unblind} and DT \cite{zhang2023self} on color images from the BSD300 \cite{roth2005fields} and Kodak \cite{franzen1999kodak} datasets. As can be seen in Fig. \ref{fig:figure color_bsd300}-\ref{fig:figure color_kodak}, N2N achieves the best denoising performance on both datasets. The image from the BSD300 dataset contains more detailed textures, resulting in lower denoising performance for all models compared to the Kodak dataset. However, even the iterative-based method IDR achieves denoising performance similar to N2N on the Kodak dataset.

\subsection{Real-world image denoising in sRGB images}

\begin{table*}[t]
	\setlength{\abovecaptionskip}{0pt}
	\setlength{\belowcaptionskip}{0pt}
	\caption{{\bf Quantitative comparison of real-world sRGB image denoising on SIDD \cite{abdelhamed2018high} and DND \cite{plotz2017benchmarking} benchmark datasets.} We get the official evaluation results from SIDD and DND benchmark websites. $\diamond$ indicates that we have retrained the model, uploaded the test results and received the results. {\bf R} indicates that the result is reported by R2R \cite{pang2021recorrupted}. $\ast$ denotes the method with self-ensemble strategy \cite{lim2017enhanced}. The highest value is highlighted in \textbf {bold}, the second is \underline{underlined}}
	\centering
	\begin{tabular*}{\hsize}{c@{\extracolsep{\fill}}l@{\extracolsep{\fill}}l@{\extracolsep{\fill}}l@{\extracolsep{\fill}}l@{\extracolsep{\fill}}l@{\extracolsep{\fill}}}
		\hline
		\multirow{2}*{Training data} & \multirow{2}*{Method} & \multicolumn{2}{c}{SIDD benchmark} & \multicolumn{2}{c}{DND} \\
		&   & PSNR & SSIM &PSNR  &SSIM \\
		\hline
		\multirow{6}*{Need noise images and other priors}&	GCBD \cite{chen2018image}&	-&	-&	35.58&	0.922\\
		&UIDNet \cite{hong2020end}&	32.48&	0.897&	-&	-\\
		&D-BSN \cite{wu2020unpaired} + MWCNN \cite{liu2018multi} 	&- 	&- 	&37.93	&bf 0.937\\
		&NAC \cite{xu2020noisy} 	          &-	&-	&36.20	&0.925\\
		&R2R \cite{pang2021recorrupted}   	&34.78	&0.898	&-	&-\\
		&C2N \cite{jang2021c2n} + DIDN\textsuperscript{$\ast$} \cite{yu2019deep} 	&35.35	&{\bf 0.937}	&37.28	&0.924\\
		\hline
		\multirow{5}*{\makecell[c]{Need single noisy}}&Noise2Void \cite{krull2019noise2void} &	27.68\textsuperscript{\bf R}	&0.668\textsuperscript{\bf R}	&-	&-\\
		&Noise2Self \cite{batson2019noise2self} 	&29.56\textsuperscript{\bf R}	&0.808\textsuperscript{\bf R}	&-	&-\\
		
		&CVF-SID (S2) \cite{neshatavar2022cvf}	&34.71	&0.917	&36.50	&0.924\\
		&AP-BSN \cite{lee2022ap}	&35.97	&0.925	&38.09	&0.937\\
		&Li et al. \cite{vaksman2023patch}	&{\bf 37.41}	&0.934	&38.18&	0.938\\
		&LG-BPN \cite{wang2023lg}&	37.28 	&\underline{0.936}	&\underline{38.43}	&\underline{0.942}\\
		&MM-BSN\cite{zhang2023mm}	     &\underline{37.37}	& \underline{0.936}	&{\bf 38.74}	&{\bf 0.943}\\
		\hline
	\end{tabular*}
	\setlength{\belowcaptionskip}{-0.2cm}
	\label{tab:srgb_sidd_dnd}
\vspace{-2em}
\end{table*}

To compare the effectiveness of self-supervised denoising models on real-world sRGB images, we evaluated them on sRGB of SIDD benchmark \cite{abdelhamed2018high}, DND benchmark \cite{plotz2017benchmarking}, CC \cite{nam2016holistic} and PolyU \cite{xu2018real} datasets using PSNR and SSIM as the metrics.

Table \ref{tab:srgb_sidd_dnd} compares the performance of several models on real-world sRGB images by SIDD \cite{abdelhamed2018high} and DND \cite{plotz2017benchmarking} benchmark datasets, where C2N \cite{jang2021c2n}, Li et al. \cite{li2023spatially}, LG-BPN \cite{wang2023lg} and MM-BSN \cite{zhang2023mm} achieving better performance. Notably, the latter three models are BSN-based methods that only require single noise images.

\begin{figure*}[t]
	\setlength{\abovecaptionskip}{0cm}
	\setlength{\belowcaptionskip}{0cm}
	\centering
	\begin{minipage}[b]{0.21\linewidth}
		\setlength{\abovecaptionskip}{1pt}
		\centering
		\captionsetup[subfloat]{labelsep=none,format=plain,labelformat=empty,font={scriptsize}}
		\subfloat[SIDD validation]
		{\includegraphics[width=1.52in]{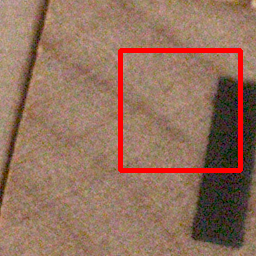}}
	\end{minipage}
	\begin{minipage}[b]{0.085\linewidth}
		\setlength{\abovecaptionskip}{1pt}
		\centering
		\captionsetup[subfloat]{labelsep=none,format=plain,labelformat=empty,font={scriptsize}}
		\subfloat[Clean]
		{\includegraphics[width=0.61in]{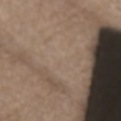}}
		\quad
		\captionsetup[subfloat]{labelsep=none,format=plain,labelformat=empty,font={scriptsize}}
		\subfloat[C2N \cite{jang2021c2n}]
		{\includegraphics[width=0.61in]{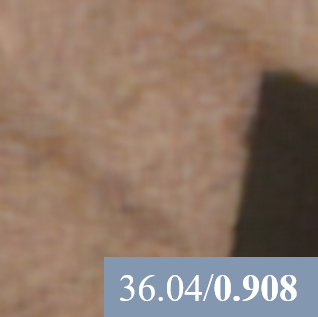}}
	\end{minipage}
	\begin{minipage}[b]{0.085\linewidth}
		\setlength{\abovecaptionskip}{1pt}
		\centering
		\captionsetup[subfloat]{labelsep=none,format=plain,labelformat=empty,font={scriptsize}}
		\subfloat[CVF-SID\cite{neshatavar2022cvf}]
		{\includegraphics[width=0.61in]{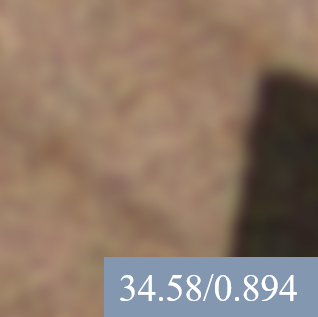}}
		\quad
		\captionsetup[subfloat]{labelsep=none,format=plain,labelformat=empty,font={scriptsize}}
		\subfloat[AP-BSN\cite{lee2022ap}]
		{\includegraphics[width=0.61in]{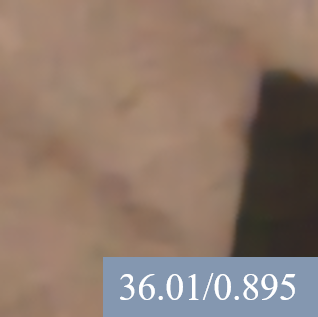}}
	\end{minipage}
	\begin{minipage}[b]{0.085\linewidth}
		\setlength{\abovecaptionskip}{1pt}
		\centering
		\captionsetup[subfloat]{labelsep=none,format=plain,labelformat=empty,font={scriptsize}}
		\subfloat[LG-BSN \cite{wang2023lg}]
		{\includegraphics[width=0.61in]{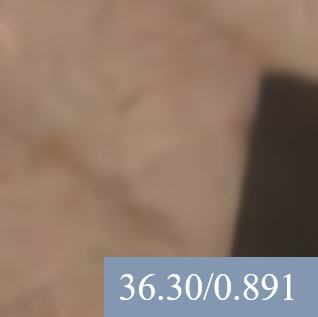}}
		\quad
		\captionsetup[subfloat]{labelsep=none,format=plain,labelformat=empty,font={scriptsize}}
		\subfloat[MM-BSN\cite{zhang2023mm}]
		{\includegraphics[width=0.61in]{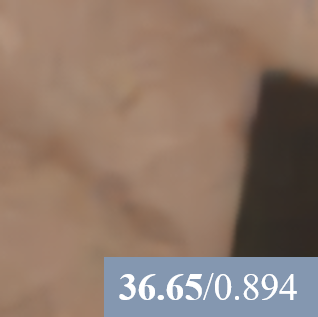}}
	\end{minipage}
	\begin{minipage}[b]{0.21\linewidth}
		\setlength{\abovecaptionskip}{1pt}
		\centering
		\captionsetup[subfloat]{labelsep=none,format=plain,labelformat=empty,font={scriptsize}}
		\subfloat[SIDD Benchmark]
		{\includegraphics[width=1.52in]{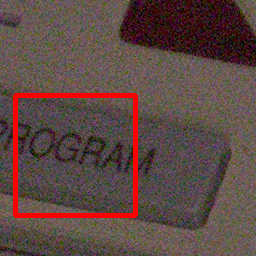}}
	\end{minipage}
	\begin{minipage}[b]{0.085\linewidth}
		\setlength{\abovecaptionskip}{1pt}
		\centering
		\captionsetup[subfloat]{labelsep=none,format=plain,labelformat=empty,font={scriptsize}}
		\subfloat[Noisy]
		{\includegraphics[width=0.61in]{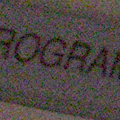}}
		\quad
		\captionsetup[subfloat]{labelsep=none,format=plain,labelformat=empty,font={scriptsize}}
		\subfloat[C2N \cite{jang2021c2n}]
		{\includegraphics[width=0.61in]{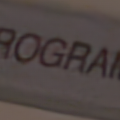}}
	\end{minipage}
	\begin{minipage}[b]{0.085\linewidth}
		\setlength{\abovecaptionskip}{1pt}
		\centering
		\captionsetup[subfloat]{labelsep=none,format=plain,labelformat=empty,font={scriptsize}}
		\subfloat[CVF-SID\cite{neshatavar2022cvf}]
		{\includegraphics[width=0.61in]{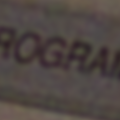}}
		\quad
		\captionsetup[subfloat]{labelsep=none,format=plain,labelformat=empty,font={scriptsize}}
		\subfloat[AP-BSN\cite{lee2022ap}]
		{\includegraphics[width=0.61in]{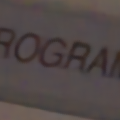}}
	\end{minipage}
	\begin{minipage}[b]{0.085\linewidth}
		\setlength{\abovecaptionskip}{1pt}
		\centering
		\captionsetup[subfloat]{labelsep=none,format=plain,labelformat=empty,font={scriptsize}}
		\subfloat[LG-BSN\cite{wang2023lg}]
		{\includegraphics[width=0.61in]{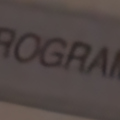}}
		\quad
		\captionsetup[subfloat]{labelsep=none,format=plain,labelformat=empty,font={scriptsize}}
		\subfloat[MM-BSN\cite{zhang2023mm}]
		{\includegraphics[width=0.61in]{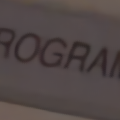}}
	\end{minipage}
	\setlength{\belowcaptionskip}{-0.5cm}
	\caption{Visual comparison of different models denoising real-world sRGB images from SIDD \cite{abdelhamed2018high} validation and benchmark datasets. }
	\label{fig:figure 10}
\vspace{-1.5em}
\end{figure*}

Table \ref{tab:srgb_cc} compares the denoising performance of N2N \cite{lehtinen2018noise2noise}, N2V \cite{krull2019noise2void}, DIP \cite{ulyanov2018deep}, N2S \cite{batson2019noise2self}, S2S \cite{quan2020self2self}, DBSN \cite{wu2020unpaired}, NAC \cite{xu2020noisy} and R2R \cite{pang2021recorrupted} on the CC \cite{nam2016holistic} dataset. Among them, S2S, DBSN, NAC, and R2R outperform N2N, demonstrating that it is possible to achieve better denoising results using single images and noise prior or unpaired noisy-clean images.

\begin{figure}[t]
	\setlength{\abovecaptionskip}{0cm}
	\setlength{\belowcaptionskip}{0cm}
	\centering
	\begin{minipage}[b]{0.52\linewidth}
		\setlength{\abovecaptionskip}{1pt}
		\captionsetup[subfloat]{labelsep=none,format=plain,labelformat=empty,font={scriptsize}}
		\subfloat[Noisy]
		{\includegraphics[width=1.803in]{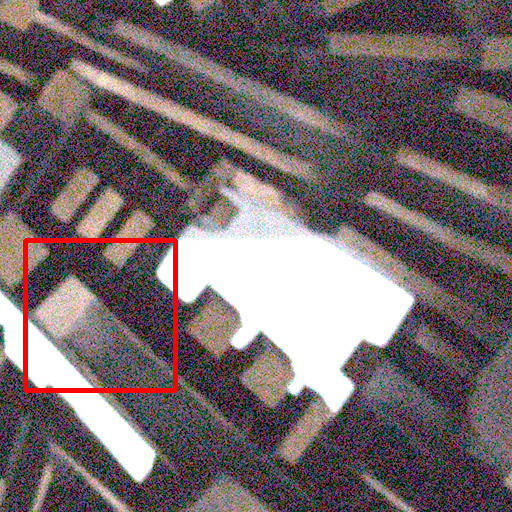}}
	\end{minipage}
	\begin{minipage}[b]{0.21\linewidth}
		\setlength{\abovecaptionskip}{1pt}
		\centering
		\addtocounter{subfigure}{-1}
		\captionsetup{font={scriptsize}}
		\subfloat[C2N\cite{jang2021c2n}]
		{\includegraphics[width=0.75in]{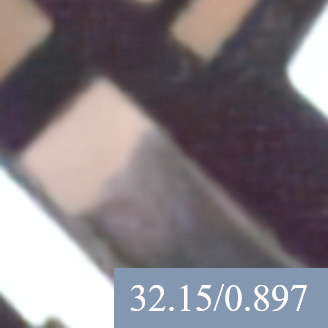}}
		\quad
		\captionsetup{font={scriptsize}}
		\subfloat[CVF-SID\cite{neshatavar2022cvf}]
		{\includegraphics[width=0.75in]{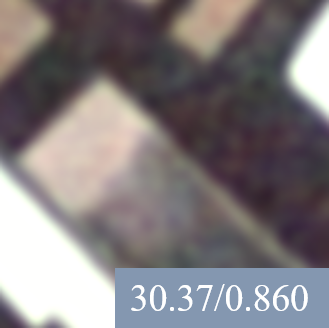}}
	\end{minipage}
	\begin{minipage}[b]{0.21\linewidth}
		\setlength{\abovecaptionskip}{1pt}
		\centering
		\captionsetup{font={scriptsize}}
		\subfloat[AP-BSN\cite{lee2022ap}]
		{\includegraphics[width=0.75in]{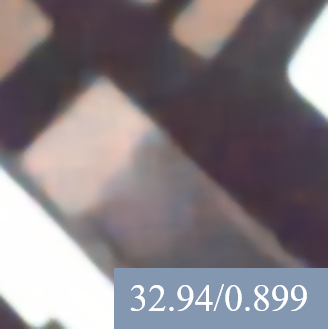}}
		\quad
		\captionsetup{font={scriptsize}}
		\subfloat[MM-BSN\cite{zhang2023mm}]
		{\includegraphics[width=0.75in]{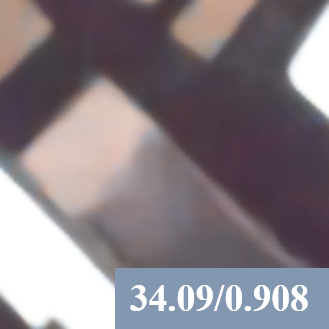}}
	\end{minipage}
	\setlength{\belowcaptionskip}{-0.5cm}
	\caption{Visual comparison of different models denoising real-world sRGB images on DND \cite{plotz2017benchmarking} benchmark datasets. }
	\label{fig:figure 11}
\vspace{-1.5em}
\end{figure}

\begin{figure*}[htbp]
	\setlength{\abovecaptionskip}{0cm}
	\centering
	\subfloat[Clean]{
	\begin{minipage}[b]{0.185\linewidth}
			\includegraphics[width=1.35in]{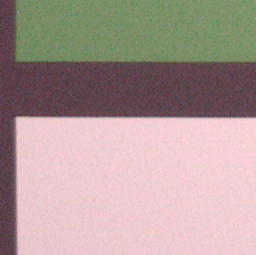}\vspace{2pt}
			\includegraphics[width=1.35in]{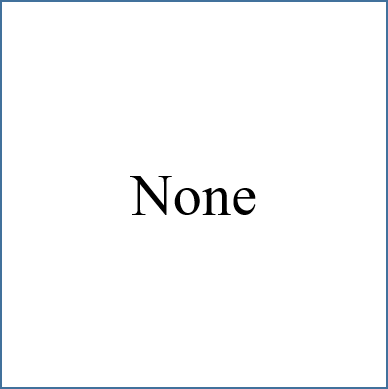}
	\end{minipage}
}
	\subfloat[Noisy]{
	\begin{minipage}[b]{0.185\linewidth}
		\includegraphics[width=1.349in]{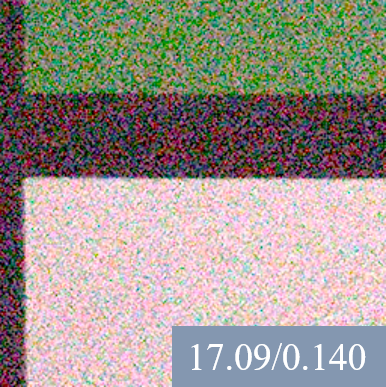}\vspace{2pt}
		\includegraphics[height=1.349in]{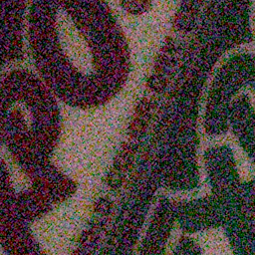}
	\end{minipage}
}
	\subfloat[NBR2NBR \cite{huang2021neighbor2neighbor}]{
	\begin{minipage}[b]{0.185\linewidth}
		\includegraphics[width=1.35in]{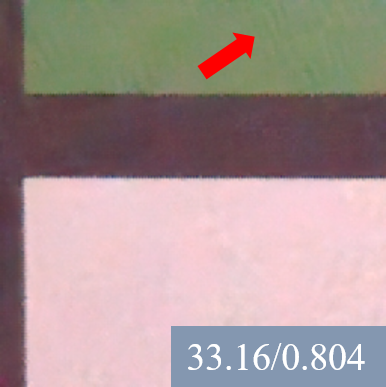}\vspace{2pt}
		\includegraphics[width=1.35in]{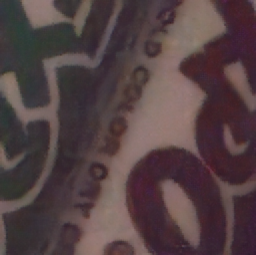}
	\end{minipage}
}
	\subfloat[B2UB \cite{wang2022blind2unblind}]{
	\begin{minipage}[b]{0.185\linewidth}
		\includegraphics[width=1.35in]{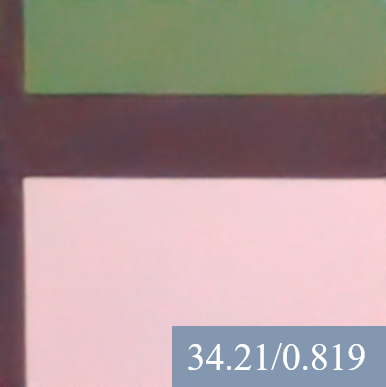}\vspace{2pt}
		\includegraphics[width=1.35in]{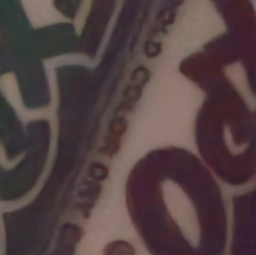}
	\end{minipage}
}	
	\subfloat[DT \cite{zhang2023self}]{
	\begin{minipage}[b]{0.185\linewidth}
		\includegraphics[width=1.35in]{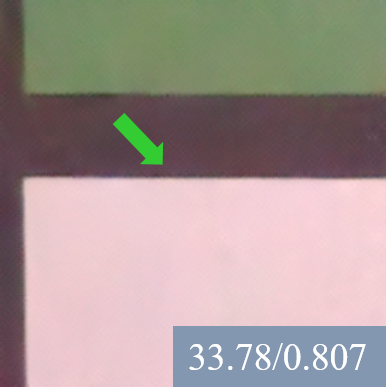}\vspace{2pt}
		\includegraphics[width=1.35in]{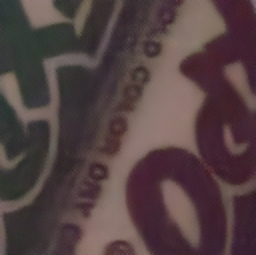}
	\end{minipage}
}

\setlength{\belowcaptionskip}{-0.5cm}
\caption{Visual Comparison of different methods denoising rawRGB images in SIDD \cite{abdelhamed2018high} validation and benchmark datasets. The conversion from raw-RGB to sRGB is performed using the official ISP tool \cite{abdelhamed2018high}.}
\label{fig:figure 12}
\vspace{-1em}
\end{figure*}

\begin{table*}
	\setlength{\abovecaptionskip}{0cm}
	\setlength{\belowcaptionskip}{0cm}
	\caption{Quantitative comparison of real-world sRGB image denoising on CC \cite{nam2016holistic} dataset.}
	\centering
	\begin{tabular*}{\hsize}{c@{\extracolsep{\fill}}c@{\extracolsep{\fill}}c@{\extracolsep{\fill}}c@{\extracolsep{\fill}}c@{\extracolsep{\fill}}c@{\extracolsep{\fill}}c@{\extracolsep{\fill}}c@{\extracolsep{\fill}}c@{\extracolsep{\fill}}}
		\hline
		Metrics&N2N \cite{lehtinen2018noise2noise}&N2V \cite{krull2019noise2void}&DIP\cite{ulyanov2018deep}&N2S \cite{batson2019noise2self}&S2S \cite{quan2020self2self}&DBSN \cite{wu2020unpaired}&NAC \cite{xu2020noisy}&R2R \cite{pang2021recorrupted}\\
		\hline
		PSNR&	35.32&	32.27&	35.69&	33.38&	\underline{37.52}&	35.90&	36.59&	{\bf 37.78}\\
		SSIM&	0.916&	0.862&	0.926&	0.846&	\underline{0.947}&	0.937&	{\bf 0.950}&	0.945\\
		\hline
	\end{tabular*}
	\setlength{\belowcaptionskip}{-0.5cm}
	\label{tab:srgb_cc}
\vspace{-2em}
\end{table*}

Table \ref{tab:srgb_polyu} compares the denoising performance of DIP \cite{ulyanov2018deep}, N2V \cite{krull2019noise2void}, N2S \cite{batson2019noise2self}, S2S \cite{quan2020self2self} and R2R \cite{pang2021recorrupted} on the PolyU \cite{xu2018real} dataset, with R2R achieving the best denoising performance, followed by S2S.

\begin{table}[t]
	\setlength{\abovecaptionskip}{0cm}
	\setlength{\belowcaptionskip}{0cm}
	\caption{Quantitative comparison of real-world sRGB image denoising on PolyU \cite{xu2018real} dataset.}
	\centering
	\begin{tabular*}{\hsize}{c@{\extracolsep{\fill}}c@{\extracolsep{\fill}}c@{\extracolsep{\fill}}c@{\extracolsep{\fill}}c@{\extracolsep{\fill}}c@{\extracolsep{\fill}}}
		\hline
		Metrics&	DIP \cite{ulyanov2018deep}&	N2V \cite{krull2019noise2void}&	N2S \cite{batson2019noise2self}&	S2S \cite{quan2020self2self}&	R2R \cite{pang2021recorrupted}\\
		\hline
		PSNR&	36.95&	34.08&	35.46&	37.52&	38.47\\
		SSIM&	0.975&	0.954&	0.965&	0.983&	0.965\\
		\hline
	\end{tabular*}
	\label{tab:srgb_polyu}
\vspace{-2em}
\end{table}

The visual denoising performance of several self-supervised models of C2N \cite{jang2021c2n}, CVF-SID \cite{neshatavar2022cvf}, AP-BSN \cite{lee2022ap}, LG-BPN \cite{wang2023lg}, and MM-BSN \cite{zhang2023mm} are compared on the SIDD validation and benchmark, as well as the DND dataset. C2N requires unpaired noisy-clean image pairs, while the remaining methods, including BSN-based methods AP-BSN, LG-BPN, and MM-BSN, and the general method CVF-SID, only need single noise images.

Fig. \ref{fig:figure 10} shows the visual comparison of five denoising methods on the SIDD\cite{abdelhamed2018high} validation and benchmark datasets. From the SIDD validation image, it can be observed that the BSN-based methods produce smoother denoised results with better noise removal, but suffer from significant loss of texture information. C2N \cite{jang2021c2n} and CVF-SID \cite{neshatavar2022cvf} preserve the detailed textures better, but CVF-SID has a poorer denoising performance. Therefore, C2N is more suitable for denoising noisy images with more detailed textures.
As shown in Fig. \ref{fig:figure 10} and Figure \ref{fig:figure 11} for the SIDD benchmark images and the DND benchmark image, the BSN-based methods achieve better denoising results. For images with higher noise level and less fine textures in the image signal, the BSN-based methods are a recommended choice.

\subsection{Real-world image denoising in rawRGB images}

Table \ref{tab:rawrgb} presents a quantitative comparison of self-supervised denoising mehods on real-world rawRGB images from the SIDD\cite{abdelhamed2018high} benchmark and validation datasets. The table shows that B2UB \cite{wang2022blind2unblind} achieves the highest PSNR and SSIM on both datasets, followed by DT \cite{zhang2023self}. Notably, both B2UB and DT are also BSN-based methods that require only single noisy images for training. DT, which combines Transformer to supplement global information, is also among the top denoisers.

\begin{table}[t]
	\setlength{\abovecaptionskip}{0cm}
	\setlength{\belowcaptionskip}{0cm}
	\caption{Quantitative comparison of real-world rawRGB image denoising on SIDD \cite{abdelhamed2018high} benchmark and validation datasets. The highest value is highlighted in \textbf {bold}, the second is \underline{underlined}}
	\centering
	\begin{tabular*}{\hsize}{c@{\extracolsep{\fill}}c@{\extracolsep{\fill}}c@{\extracolsep{\fill}}c@{\extracolsep{\fill}}c@{\extracolsep{\fill}}}
		\hline
		\multirow{2}*{Method}&	\multicolumn{2}{c}{SIDD benchmark}&	\multicolumn{2}{c}{SIDD validation}\\
		&PSNR	&SSIM	&PSNR	&SSIM\\
		\hline
		N2V \cite{krull2019noise2void}	&48.01	&0.983	&48.55	&0.984\\
		Laine19-mu (G) \cite{laine2019high}&	49.82&	0.989&	50.44&	0.990\\
		Laine19-pme (G)\cite{laine2019high}&	42.17&	0.935&	42.87&	0.939\\
		Laine19-mu (P)\cite{laine2019high}&	50.28&	0.989&	50.89&	0.990\\
		Laine19-pme (P)\cite{laine2019high}&	48.46&	0.984&	48.98&	0.985\\
		DBSN\cite{wu2020unpaired}&	49.56&	0.987&	50.13&	0.988\\
		R2R\cite{pang2021recorrupted}&	46.70&	0.978&	47.20&	0.980\\
		NBR2NBR\cite{huang2021neighbor2neighbor}&	50.47&	0.990&	51.06&	0.991\\
		B2UB\cite{wang2022blind2unblind}&	{\bf 50.79}&	{\bf 0.991}&	{\bf 51.36}&	{\bf 0.992}\\
		DT\cite{zhang2023self}&	\underline{50.62}&	\underline{0.990}&	\underline{51.16}&	\underline{0.991}\\
		\hline
	\end{tabular*}
	\label{tab:rawrgb}
\vspace{-2em}
\end{table}

Fig. \ref{fig:figure 12} depicts denoising results for various methods on the SIDD\cite{abdelhamed2018high} validation and benchmark datasets. The first row shows images from the SIDD validation, while the second row shows images from the SIDD benchmark. As shown in the figure, the denoising performance of NBR2NBR \cite{huang2021neighbor2neighbor} is noticeably worse than that of the other methods, especially on the SIDD validation image. The denoised image contains artifacts, as indicated by the red arrow in the figure. The BSN-based methods B2UB \cite{wang2022blind2unblind} and DT \cite{zhang2023self} that combines BSN and Transformer, have similar noise removal capabilities. However, as shown by the green arrow in the figure, DT preserves the details of the image signal better.

\section{Future direction}
\label{sec:Section5}
In recent years, self-supervised denoising methods based on deep neural networks have gained increasing attention. These methods have demonstrated comparable denoising performance to supervised methods and show great potential for further improvement. In this section, we summarize some of the challenges and directions in the self-supervised image denoising field based on survey results:

1.	Most existing self-supervised denoising methods that require only single noisy images for model training are based on BSN and achieves the state-of-the-art performance. While these methods achieve good denoising results under the assumption of independent noise and zero-mean, real-world sRGB images often have spatially correlated noise that violates this assumption. Although some methods \cite{lee2022ap, zhang2023mm, wang2023lg} have been proposed to destroy the spatial correlation of noise, they also destroy the texture information of the image signal. Therefore, developing methods that can preserve the detailed texture of the image signal while destroying the spatial connection of the noise would greatly improve the performance of all BSN-based models for denoising real sRGB images.

2.	Existing self-supervised models are established based on small noise structures such as Gaussian noise, Poisson noise, or real noise in SIDD images. They cannot handle special noises like stripe noise or speckle noise. Therefore, image denoising with large areas of spatially correlated noise is a challenging direction.

3.	CNNs have limitations in controlling global information due to the restricted receptive field. Transformers can help address this limitation by providing a global view of the image. However, attempts to combine CNN and Transformer for image denoising, such as in DT \cite{zhang2023self} and LG-BPN \cite{wang2023lg}, did not achieve the expected improvement of "one plus one bigger than two". Given that the combination of CNN and Transformer is still in the early stages of research, there is still room for improvement in this area, making it a promising direction for future research. Developing more effective ways to integrate CNN and Transformer architectures for image denoising could lead to significant improvements in denoising performance.

4.	Artificial Intelligence Generated Content (AIGC) is gaining more attention with the application of diffusion denoising methods\cite{fadnavis2020patch2self,yang2023real,kong2023comparison}, which have achived outstanding performance. We believe that diffusion models have great potential in self-supervised denoising, and it is our next reseach direction.

\section{Conclusion}
\label{sec:Section6}
This paper provides a comprehensive survey of recent self-supervised image denoising models, dividing them into three categories: General methods, BSN-based methods, and Transformer-based methods. To the best of our knowledge, this is the first survey to focus solely on self-supervised image denoising methods. The paper provides a brief introduction to classical methods in each of the three categories and evaluates them quantitatively and qualitatively on various datasets.

From a large number of experimental results, the following findings were observed: (1) For the remaining general datasets, BSN-based methods perform better than other methods on most datasets. (2) BSN-based methods are prone to destroy texture information in image signals when breaking the noise spatial connection, leading to smoother denoised images with fine textures lost. Thus, for noisy images with more fine textures, general methods are more recommended. (3) The self-supervised Transformer-based image denoising method achieves the comparable denoising performance, but has not surpassed the performance of using CNN alone. Based on the above analyses, we tend to believe that future image noise reduction research should take the BSN-BASED method as the benchmark, and explore new ideas for large-area spatial contact noise reduction based on the Transformer and Diffusion models, in order to make the noise reduction theory more relevant to practical applications.

{\small
	\bibliographystyle{ieee_fullname}
	\bibliography{my_ref}
}

\end{document}